\pdfoutput=1
\documentclass[10pt]{article}
\usepackage[utf8]{inputenc}
\usepackage{cite}
\usepackage{hyperref}
\hypersetup{colorlinks=true,linkcolor=bluecyan,citecolor=orange3,urlcolor=green3,pdfencoding=auto,linktocpage}
\usepackage{amsmath,amssymb,amsbsy,amstext,amsthm,simplewick,amsfonts,wasysym}
\usepackage{caption}
\usepackage{mathrsfs}
\usepackage{graphicx}
\usepackage{wrapfig}
\usepackage{upgreek}
\usepackage{bm} 
\usepackage{framed}
\usepackage{bbm}
\usepackage{textcomp}
\usepackage{adjustbox}
\usepackage{makecell}
\usepackage{tcolorbox}
\usepackage{empheq}
\usepackage[normalem]{ulem}
\usepackage{enumitem}
\usepackage{braket} 
\usepackage{array}
\usepackage{dsfont}
\usepackage{physics}
\usepackage{ulem}
\usepackage{xcolor}
\usepackage{caption}
\usepackage{subcaption}
\usepackage{tocloft}
\usepackage{tikz}
\usepackage{relsize}
\usepackage{overpic}

\usetikzlibrary{calc}
\usetikzlibrary{decorations.markings}
\usetikzlibrary{decorations.pathmorphing}
\usetikzlibrary{shapes.misc}
\usetikzlibrary{arrows}
\usetikzlibrary{shapes.geometric}
\usetikzlibrary{plotmarks}

\usepackage[framemethod=default]{mdframed}

\newmdenv[backgroundcolor=gray!15,%
skipabove=5pt,%
skipbelow=5pt,%
leftmargin=2pt,%
rightmargin=2pt,%
innertopmargin=-6pt,%
innerbottommargin=5pt,%
innerleftmargin=5pt,%
innerrightmargin=5pt,%
splittopskip=0pt,%
splitbottomskip=0pt,%
linewidth=0pt,%
nobreak=true]%
{keyeqn}

\newmdenv[backgroundcolor=gray!13,%
skipabove=5pt,%
skipbelow=5pt,%
leftmargin=0pt,%
rightmargin=0pt,%
innertopmargin=-2pt,%
innerbottommargin=5pt,%
innerleftmargin=5pt,%
innerrightmargin=5pt,%
splittopskip=0pt,%
splitbottomskip=0pt,%
linewidth=0pt,%
nobreak=true]%
{keythrm}

\graphicspath{{Fig/}}

\definecolor{lightgreen}{cmyk}{0.2, 0, 0.2, 0.2}
\definecolor{lightgray}{cmyk}{0.1,0.2,0,0.1}
\definecolor{lightgray2}{cmyk}{0.1,0.1,0,0.1}
\definecolor{bluecyan}{RGB}{0, 100, 200}
\definecolor{blue3}{RGB}{31,119,180}
\definecolor{red3}{RGB}{214,39,40}
\definecolor{orange3}{RGB}{255,127,14}
\definecolor{green3}{RGB}{44,160,44}

\definecolor{red2}{RGB}{255,0,0}
\definecolor{green2}{RGB}{0,170,0}
\definecolor{blue2}{RGB}{0,128,255}
\definecolor{magenta2}{RGB}{191,64,191}
\definecolor{purple2}{RGB}{112,48,160}
\definecolor{orange2}{RGB}{255,192,0}

\newtheorem{lemma}{Lemma}[section]

\def\bfk{\textbf{k}}
\def\bfq{\textbf{q}}

\def\bfx{\textbf{x}}

\def\Re{\mathrm{Re}\,}
\def\Im{\mathrm{Im}\,}
\def\d{\mathrm{d}}

\numberwithin{equation}{section}

\allowdisplaybreaks[1]
\begin{document}

\begin{titlepage}
	\setcounter{page}{1} \baselineskip=15.5pt 
	\thispagestyle{empty}

     \begin{center}
		{\fontsize{18}{18}\centering {\bf{A Match Made in Heaven:}} \\ \vspace{0.2cm} {\textit{Linking Observables in    Inflationary Cosmology}}\;}\\
	\end{center}
 
	\vskip 18pt
	\begin{center}
		\noindent
		{\fontsize{12}{18}\selectfont David Stefanyszyn\footnote[1]{\tt david.stefanyszyn@nottingham.ac.uk}$^{,a}$, Xi Tong  \footnote[2]{\tt xt246@cam.ac.uk}$^{,b}$ and Yuhang Zhu \footnote[3]{\tt yhzhu@ibs.re.kr}$^{,c}$}
	\end{center}
	
	\begin{center}
		\vskip 8pt
		$a$ \textit{School of Mathematical Sciences \& School of Physics and Astronomy,
			University of Nottingham, University Park, Nottingham, NG7 2RD, UK} \\
        $b$ \textit{Department of Applied Mathematics and Theoretical Physics, University of Cambridge,\\Wilberforce Road, Cambridge, CB3 0WA, UK} \\
        $c$ \textit{Cosmology, Gravity and Astroparticle Physics Group,\\
        Center for Theoretical Physics of the Universe,\\
        Institute for Basic Science, Daejeon 34126, Korea}
	\end{center}

    \vskip 40pt
	
	\noindent\rule{\textwidth}{0.4pt}
	\noindent \textbf{Abstract} ~~ Cosmological correlation functions of inflaton and graviton perturbations are the fundamental observables of early universe cosmology and remain a primary target for observations. In this work, we ask the following question: are these observables independent of one another? We find that in the parity-odd sector of inflationary perturbation theory, the answer is a resounding no! In earlier work we derived a \textit{correlator-to-correlator} factorisation formula which states that parity-odd correlators factorise into lower-point correlators under some mild assumptions on the underlying theory. In this work, we show that these assumptions are satisfied in \textit{dynamical Chern-Simons gravity} where the action of minimal inflation is augmented by a coupling between the inflaton and the gravitational Chern-Simons term. Such a theory gives rise to a parity-odd trispectrum of curvature perturbations, and we show that such a trispectrum can be expressed solely in terms of the bispectrum that arises due to the minimal coupling between the inflaton and graviton, and the graviton power spectrum which receives a parity-odd correction in this theory. The trispectrum is quadratic in this mixed inflaton-graviton bispectrum and can therefore be interpreted as a ``double copy". Our final expression for the parity-odd trispectrum is a relatively simple function of the external momenta that is rational and factorised.        

	
	
\end{titlepage} 


\newpage
\setcounter{tocdepth}{2}
\setcounter{page}{2}
{   \noindent\rule{\textwidth}{0.4pt}
	\tableofcontents
    \noindent\rule{\textwidth}{0.4pt}
}


\section{Introduction}

Detecting the primordial power spectrum of gravitons, and non-Gaussianities of inflatons and gravitons, remains a primary focus of early universe cosmology. Generically, such observables are viewed as independent quantities with each containing its own crucial information on the inflationary dynamics. From the graviton power spectrum we can learn about the energy scale of inflation \cite{Lyth:1996im}, while through non-Gaussianities we can learn about inflationary interactions \cite{Maldacena:2002vr,Chen:2006nt,Cheung:2007st} and even see evidence of new degrees of freedom \cite{Chen:2009zp,Baumann:2011nk,Noumi:2012vr,Arkani-Hamed:2015bza,Lee:2016vti}. In traditional approaches such as the in-in/Schwinger-Keldysh formalisms \cite{Weinberg:2005vy,Chen:2017ryl}, different correlators are computed individually in perturbation theory via Feynman diagram expansions. A drawback of this approach is that individual diagrams are not physical observables and are technically challenging to compute given the presence of nested time integrals. This has prompted much effort to develop new computational techniques \cite{Benincasa:2022omn,Benincasa:2022gtd,Arkani-Hamed:2017fdk,Albayrak:2023hie,Benincasa:2024leu,Arkani-Hamed:2018kmz,Baumann:2019oyu,Baumann:2020dch,Baumann:2021fxj,Arkani-Hamed:2023kig,Pajer:2020wxk,Jazayeri:2021fvk,Bonifacio:2021azc,Salcedo:2022aal,Donath:2024utn,Melville:2023kgd,Pimentel:2022fsc,Jazayeri:2022kjy,Chowdhury:2023arc,Gomez:2021qfd,Hogervorst:2021uvp,DuasoPueyo:2023kyh,Aoki:2024uyi,Chowdhury:2025ohm,Armstrong:2022mfr,Mei:2024abu,DiPietro:2021sjt,Tong:2021wai,Qin:2022fbv,Xianyu:2022jwk,Werth:2023pfl,Aoki:2023wdc,Xianyu:2023ytd,Qin:2023bjk,Werth:2024mjg,Ema:2024hkj,Liu:2024xyi,Liu:2024str,Grimm:2025zhv,Cespedes:2025dnq,Bzowski:2013sza,McFadden:2011kk,Bzowski:2011ab,Sleight:2019hfp,Sleight:2019mgd,Sleight:2021plv}. Interestingly, these mathematical complexities can sometimes cancel when diagrams are summed together, leaving a structurally simple final correlator. 

Examples of this behaviour are ever-present in the parity-odd (PO) sector of primordial perturbations \cite{Cabass:2021fnw, Liu:2019fag, Cabass:2022rhr, Thavanesan:2025kyc, Goodhew:2024eup,Jazayeri:2023kji, Stefanyszyn:2023qov,Lee:2023jby}. Within this sector, certain kinematic singularities that are characteristic of cosmological observables, and which indicate that one has had to work hard to arrive at the final result, drop out once all contributions are summed together. Furthermore, the final result can be non-linearly related to other lower-point cosmological correlators \cite{Stefanyszyn:2024msm}. Such relations hold for general physical kinematics and are therefore distinct from cosmological cutting rules \cite{Goodhew:2020hob,Melville:2021lst,Goodhew:2021oqg,Meltzer:2021zin,Cespedes:2020xqq} and soft theorems/consistency relations \cite{Creminelli:2012ed,Creminelli:2013mca,Hinterbichler:2013dpa,Maldacena:2002vr}.

More specifically, in \cite{Stefanyszyn:2024msm} we showed that PO correlators of curvature perturbations $\zeta$ (which ultimately seed the temperature anisotropies of the CMB and density perturbations of large scale structure) factorise into lower-point correlators. As an example, consider a PO four-point function $B_{\zeta \zeta \zeta \zeta}^{\text{PO}}$ that arises due to the exchange of a new spinning degree of freedom $\sigma$ that couples to $\zeta$ via both a parity-even and a parity-odd cubic coupling. In such a theory, the resulting PO four-point function can be written in terms of the two-point function of the new field $B_{\sigma \sigma}$ and the three-point function that mixes the two states $B_{\zeta \zeta \sigma}$ \cite{Stefanyszyn:2024msm} (with indices suppressed):
\begin{align}
B^{\text{PO}}_{\zeta \zeta \zeta \zeta} = \left[ B_{\zeta \zeta\sigma} \cdot \frac{1}{B_{\sigma \sigma}} \cdot B_{\zeta \zeta \sigma} \right]^{\text{PO}} \,,
\end{align}
where $\cdot$ denotes that indices are contracted (or helicities are summed over) and $[\ldots]^{\text{PO}}$ denotes that we project onto the PO part. 
We refer to such a factorisation as \textit{correlator-to-correlator factorisation} (CCF). This is a very non-trivial relation between different cosmological observables and holds under some very mild assumptions:
\begin{itemize}
  \item  Unitarity and locality \hspace{3cm} \,\textbullet\ Tree-level approximation
  \item Bunch-Davies vacuum     \hspace{3cm} \textbullet\ Complementary/discrete series mass for $\sigma$
  \item Scale invariance and the absence of late-time divergences
\end{itemize}
In \cite{Stefanyszyn:2024msm,Stefanyszyn:2023qov} we considered a number of examples where $\sigma$ corresponds to an isocurvature perturbation. However, a more tantalising prospect is where the ``new" degree of freedom is the graviton i.e. $\sigma=\gamma$. Such a scenario is in principle possible given the above listed assumptions since the graviton is massless and therefore falls within the discrete series of de Sitter representations (see \cite{Sun:2021thf} for a review). In this paper we present a concrete model whereby such a scenario is realised. We will show that the resulting PO scalar trispectrum is a ``\textit{double-copy}'' of the scalar-graviton mixed bispectra,
\begin{keyeqn}
\begin{align} \label{CCFgraviton}
B^{\text{PO}}_{\zeta \zeta \zeta \zeta} =  \left[B_{\zeta \zeta \gamma} \cdot \frac{1}{B_{\gamma \gamma}} \cdot B_{\zeta \zeta \gamma}\right]^{\rm PO}  \,,
\end{align}
\end{keyeqn}
where $B_{\gamma \gamma}$ is the graviton power spectrum, and $B_{\zeta \zeta \gamma}$ is the bispectrum that arises from the familiar minimal coupling of the inflaton to gravity first computed by Maldacena \cite{Maldacena:2002vr}.

The model that allows us realise this \textit{match made in heaven} is the dynamical Chern-Simons theory where the usual action of vanilla single-field inflation is augmented by a parity-odd coupling $\phi R_{\mu\nu\rho \sigma} \widetilde{R}^{\mu \nu \rho \sigma}$ where $R_{\mu \nu \rho \sigma}$ is the Riemann tensor and $\widetilde{R}^{\mu \nu \rho \sigma}$ is its Hodge dual. This is the first gauge-invariant parity-odd coupling between the inflaton and graviton that appears in the derivative expansion. In order to keep calculational control, we treat this correction perturbatively. Once the non-dynamical modes of the metric have been integrated out, this new operator has three primary effects on the theory of inflationary fluctuations: I) the generation of a PO correction to the graviton quadratic action, II) the generation of a PO cubic interaction of the form $\zeta \zeta \gamma$ and III) the generation of a PO inflaton quartic self-interaction.\footnote{Quartic self-interactions are familiar from computations of the parity-even trispectrum due to the coupling to gravity \cite{Seery:2006vu,Seery:2008ax,Bonifacio:2022vwa,Arroja:2008ga} and must be considered to maintain gauge invariance. The same is true here.} In principle each of these three possibilities can contribute to $B^{\text{PO}}_{\zeta \zeta \zeta \zeta}$, however using the no-go theorems of \cite{Liu:2019fag,Cabass:2022rhr,Thavanesan:2025kyc} and reality theorems of \cite{Stefanyszyn:2023qov,Goodhew:2024eup}, we will show that the latter two possibilities do not contribute: their effects on the four-point function simply drop out once we project onto the PO component. This result was also arrived at in \cite{GiorgioParity} via different methods. These theorems are relevant since the PO sector of an inflaton correlator is simply the imaginary part of the full object, and the theorems state that under certain assumptions cosmological correlators are purely real. The first contribution does indeed contribute to final PO four-point function, and does so in the factorised form in \eqref{CCFgraviton}. The fact that the final result takes this factorised form is very non-trivial, and again relies on the no-go/reality theorems. In terms of the external momenta $\bfk_1, \ldots \bfk_4$, the final form of the four-point function that we find in dynamical Chern-Simons gravity is relatively simple and takes the form:
\begin{keyeqn}
	\begin{align} 
		B_{\zeta \zeta \zeta \zeta}^{\text{PO}}  
		=&-i\left(\frac{H}{\dot{\phi}_0}\right)^4\,\frac{\pi\kappa H^7}{2M^2_{\rm{pl}}}\frac{[\hat{\bfk}_1\cdot\hat{\bfk}_3-(\hat{\bfk}_{1}\cdot\hat{\bf{s}})(\hat{\bfk}_{3}\cdot\hat{\bf{s}})]}{(k_1 k_2 k_3 k_4)^2}\frac{\hat{\bf{s}}\cdot(\hat{\bfk}_2\times\hat{\bfk}_4)}{s^3 E^2_L E^2_R} \left(E_L^3-E_L\,e_2-e_3\right) \left(E_R^3-E_R\,\tilde{e}_2-\tilde{e}_3\right)\nonumber\\
		&+\text{2 perms.,}
	\end{align}
\end{keyeqn}
where $H$ and $M_{\text{pl}}$ are respectively the Hubble scale during inflation and the Planck mass, and $\kappa$ is the Chern-Simons coupling. See the notation section below for the various definitions for the kinematic factors. Most notably, our result is a rational and factorised function of the external momenta with the absence of any \textit{total-energy} singularities. 

Our final result differs from that computed in \cite{Creque-Sarbinowski:2023wmb}. In that work, the authors considered the contribution due to the PO correction to the graviton quadratic action (contribution I in our terminology). At the level of the in-in Feynman rules, three nested time integrals need to be computed which is somewhat intractable. For this reason, in \cite{Creque-Sarbinowski:2023wmb} an approximation was made for the mode functions such that the shape is proportional to the corresponding parity-even one \cite{Seery:2008ax}. However, in this work we will show that no such approximation is necessary. Indeed, the ``complicated" parts of the computation, corresponding to nested time integrals, cancel out in the final result and therefore we do not need to compute them. This is a consequence of the no-go/reality theorems and realising that the theorems hold in dynamical Chern-Simons gravity is crucial in our ability to arrive at an exact result that does not rely on any approximations (other than the tree-level truncation). Showing that these theorems hold will form a major part of this paper. Although in this paper we will treat the Chern-Simons correction perturbatively and therefore use the usual de Sitter mode functions, this is \textit{not} why the no-go/reality theorems hold. Indeed, we have checked that if one uses the non-perturbative mode functions for the graviton, which are confluent hypergeometric functions, the theorems still hold and the final trispectrum is factorised and therefore has no total-energy poles. Our perturbative approach captures the full result at linear order in the Chern-Simons coupling, which is where we have calculational control.  

Our \textit{scalar-tensor CCF} is certainly interesting on theoretical grounds as it relates observables involving only the inflaton and graviton which are the states appearing in all inflationary models. Furthermore, we expect such relations to hold beyond the concrete model we are considering in this work. The relation is also interesting on observational grounds. Indeed, there has been much interest in recent years on cosmological parity-violation since the purported detection of mirror asymmetry in the BOSS data \cite{Philcox:2022hkh,Cabass:2022oap,Philcox:2023ffy,Philcox:2023ypl,Philcox:2024mmz,Coulton:2023oug,Hou:2022wfj,Jamieson:2024mau,Hou:2024udn,Paul:2024uim,Bao:2025onc}. A natural explanation of such parity-violation in the statistics of galaxies would be that it is of primordial origin and seeded into the initial conditions of the universe during inflation. Our scalar-tensor CCF formula therefore relates three observables each of which are of observational relevance.

The rest of the paper is organised as follows. In Section \ref{sec2} we introduce the dynamical Chern-Simons theory and outline the three different couplings that can in principle contribute to the PO four-point function of curvature perturbations. In Section \ref{sec3} we will introduce the wavefunction of the universe formalism and show how cosmological correlators, that we are ultimately interested in, are related to wavefunction coefficients. We include this section since we will use the wavefunction formalism as a means to compute cosmological correlators. This approach allows us to most easily apply the no-go/reality theorems which are crucial in our ability to compute the exact form of the tree-level four-point function in this theory. In Section \ref{sec4} we present the explicit computations and consider each of the three different couplings I,II and III. We show that the contribution from I is factorised and compute its precise form. We then show that the contributions from II and III vanish. They yield non-zero contributions to the wavefunction coefficients but drop out once we convert to the correlator. Such a phenomenon is now well-understood \cite{Maldacena:2011nz,Liu:2019fag,Cabass:2022rhr,Cabass:2021fnw,Stefanyszyn:2023qov}. In Section \ref{sec5} we discuss the scalar-graviton CCF formula in more detail and conclude in Section \ref{sec6}. In Appendix \ref{Sec:appA}, we provide an alternative derivation of our CCF expression with the traditional Schwinger-Keldysh formalism. We also provide details on how to perform the numerical integration for contribution I and compare the numerical result with our exact expression and the one derived in \cite{Creque-Sarbinowski:2023wmb}.

\paragraph{Notations and conventions}
Throughout this paper, we work with the $(-, +, +, +)$ metric sign convention, and use the following conventions for Fourier transforms
\begin{align}
    f(\bfx)=\int \frac{d^3k}{(2\pi)^3} e^{i\bfk\cdot\bfx}f(\bfk)\equiv\int_{\bfk}e^{i\bfk\cdot\bfx} f(\bfk)\,,
\end{align}
where bold letters represent three-dimensional spatial vectors. For the sake of simplicity, we also use the hatted Dirac delta function, defined as
\begin{align}
    \hat{\delta}\left(\sum \bfk\right)\equiv (2\pi)^3 \delta^{(3)} \left(\sum \bfk\right)\,.
\end{align}
The graviton field is transformed into momentum space and decomposed into the helicity basis as
\begin{align} \label{GravitonHelicityBasis}
    \gamma_{ij}(\eta,\bfx)=\int_{\mathbf{k}} e^{i\bfk\cdot\bfx} \sum_{h=\pm} {\rm{e}}^{(h)}_{ij}(\bfk) \gamma_h(\bfk,\eta)\,,
\end{align}
where $h$ is always the helicity index and the traceless and transverse polarisation tensor ${\rm{e}}^{(h)}_{ij}(\bfk)$ satisfies the following properties
\begin{align}
     \left[{\rm{e}}^{(h)}_{ij}(\bfk)\right]^*={\rm{e}}^{(h)}_{ij}(-\bfk)&={\rm{e}}^{(-h)}_{ij}(\bfk)\,, \label{PolarReality} \\
     {\rm{e}}^{(h)}_{ij}(\bfk) {\rm{e}}^{(h’)}_{ij}(-\bfk)&=2\,\delta_{hh'}\,.\label{PolarNormal}
\end{align}
The polarisation tensor can be constructed using the unit polarisation vector $\hat{\mathbf e}^{(h)}(\mathbf{\hat{k}})$ as
\begin{align}
     {\rm{e}}^{(h)}_{ij}(\bfk)=\sqrt{2}\,\hat{\rm e}^{(h)}_{i}(\mathbf{\hat{k}})\hat{\rm e}^{(h)}_{j}(\mathbf{\hat{k}})\,,
\end{align}
where 
\begin{align}
\hat{\mathbf{e}}^\pm(\hat{\mathbf{k}})=\frac{\hat{\mathbf{n}}-(\hat{\mathbf{n}}\cdot\hat{\mathbf{k}})\hat{\mathbf{k}}\pm i \,\hat{\mathbf{k}}\times\hat{\mathbf{n}}}{\sqrt{2[1-(\hat{\mathbf{n}}\cdot\hat{\mathbf{k}})^2]}}~,
\end{align}
with $\hat{\mathbf{n}}$ is an arbitrary unit vector that is not parallel to $\hat{\mathbf{k}}$. This unit vector satisfies 
\begin{align} \label{UnitVector}
[\hat{\rm e}^{\pm}_{i}(\bfk)]^* = \hat{\rm e}^{\pm}_{i}(-\bfk) =\hat{\rm e}^{\mp}_{i}(\bfk)  \,.
\end{align}
When expressing the trispectrum, we introduce Mandelstam-like variables:
\begin{align}
	\nonumber\mathbf{s}&=\mathbf{k}_1+\mathbf{k}_2~, &\mathbf{t}&=\mathbf{k}_1+\mathbf{k}_3~, &\mathbf{u}&=\mathbf{k}_1+\mathbf{k}_4~,\\
	s&=|\mathbf{k}_1+\mathbf{k}_2|~,& t&=|\mathbf{k}_1+\mathbf{k}_3|~,& u&=|\mathbf{k}_1+\mathbf{k}_4|~,
\end{align}
which satisfy the non-linear relation
\begin{align}
	k_{1}^2+k_{2}^2+k_{3}^2+k_{4}^2 = s^2+t^2+u^2,
\end{align}
with $k_{\sf a}\equiv|\bfk_{\sf{a}}|$ being the external energy variables. We also use the shorthand notations for energy variables, like $k_{\sf ab}\equiv k_{\sf a}+k_{\sf b}$. The \textit{left} and \textit{right} total energies, as well as elementary symmetric polynomials for $s$-channel kinematics are defined as 
\begin{align}
 \quad E_{L} = k_1 + k_2 + s, &\quad E_R = k_3 + k_4 +s\,, \\
e_{2} = k_1 k_2 + s(k_1+k_2), \quad e_3 = k_1 k_2 s, &\quad \tilde{e}_{2} = k_3 k_4 + s(k_3+k_4), \quad \tilde{e}_3 = k_3 k_4 s \,.
\end{align}

\section{Chern-Simons gravity and parity-odd trispectra} \label{sec2}

In this section, we briefly review inflation with dynamical Chern-Simons gravity and discuss its implication for parity violation in the trispectrum of primordial curvature perturbations (see e.g. \cite{Alexander:2009tp, Bartolo:2017szm} for other work on Chern-Simons gravity in the context of inflation). We start with the usual Lagrangian of minimal inflationary cosmology and augment it with the dynamical Chern-Simons term:
\begin{align}
    S= \int \d^4 x \sqrt{-g}\left[\frac{M^2_{\rm{pl}}}{2}R -\frac{1}{2}g^{\mu\nu}\partial_\mu \phi \partial_\nu \phi - V(\phi)-\frac{\phi}{4 f} R_{\mu\nu\rho\sigma} \widetilde{R}^{\mu\nu\rho\sigma}\right]~,\label{definingAction}
\end{align}
where
\begin{align}
    \widetilde{R}^{\mu\nu\rho\sigma}\equiv\frac{1}{2}\mathcal{E}^{\mu\nu\alpha\beta}R^{\rho\sigma}_{~~\alpha\beta} \,,
\end{align}
is the Hodge dual of the Riemann tensor and $\mathcal{E}^{\mu\nu\alpha\beta}=\epsilon^{\mu\nu\alpha\beta}/\sqrt{-g}$ is the Levi-Civita tensor density. Note that thanks to the total antisymmetry of the Levi-Civita tensor density, the Chern-Simons term can also be written in the same form using the Weyl tensor \cite{Grumiller:2007rv,Bartolo:2017szm},
\begin{align}
	R_{\mu\nu\rho\sigma} \widetilde{R}^{\mu\nu\rho\sigma}=W_{\mu\nu\rho\sigma} \widetilde{W}^{\mu\nu\rho\sigma}~,
\end{align}
where the Weyl tensor is given by
\begin{align}
	W_{\mu\nu\rho\sigma}=R_{\mu\nu\rho\sigma}-\left(g_{\mu[\rho}R_{\sigma]\nu}-g_{\nu[\rho}R_{\sigma]\mu}\right)+\frac{1}{3}g_{\mu[\rho}g_{\sigma]\nu}R~,
\end{align}
where the square brackets denote anti-symmetrisation as usual. This form of the Chern-Simons term is often useful when analysing the implications of such a correction to the inflationary action since the Weyl tensor vanishes on any conformally flat background metric. With this in mind, we can assume that the background dynamics is driven by the inflaton potential $V(\phi)$ and the Einstein-Hilbert term, leading to a slow-roll phase of inflation with\footnote{As usual, the assumptions that $\dot\phi_0\approx \text{const}$ and $H \approx \text{const}$ means that we are assuming the first slow-roll parameter $\varepsilon \approx \text{const}$. This is a valid approximation to the order we will be working since the time derivatives of $\varepsilon$ are higher-order in the slow roll expansion \cite{Bartolo:2004if}.}
\begin{align}
    \phi=\phi_0(t)\approx \dot\phi_0 t~, \quad \dot\phi_0\approx \text{const}~,\label{inflatonSlowRollBgsol}
\end{align}
and a quasi-de Sitter geometry
\begin{align}
    ds^2\approx -dt^2+a^2(t) d\mathbf{x}^2= a^2(\eta) (-d \eta^2+ d\mathbf{x}^2)~,\quad a=e^{H t}= -\frac{1}{H\eta}\,.\label{quasidSBgsol}
\end{align}
Going away from the homogeneous and isotropic background, we turn on perturbations of the inflaton field
\begin{align}
\phi=\phi_0(t)+\varphi(t,\mathbf{x})~,\label{varphiPertDef}
\end{align}
and parametrise the perturbed metric in the Arnowitt–Deser–Misner (ADM) form as
\begin{align}
	ds^2=-N^2 dt^2 +h_{ij}(d x^i+N^i dt) (d x^j+N^j dt)~,
\end{align}
where the lapse function $N$, the shift vector $N^i$ and the spatial metric $h_{ij}$ are local fields. We will work with the spatially flat gauge where the spatial metric takes the form
\begin{align}
	h_{ij}=a^2(t)(\delta_{ij}+\gamma_{ij})~,\quad \text{with}\quad \partial_i \gamma^{ij}=0~,\quad \gamma^i_{~i}=0~,\label{gammaPertDef}
\end{align}
where the indices are raised and lowered with the spatial metric $h_{ij}$. The symmetric transverse traceless tensor $\gamma_{ij}$ therefore carries two dynamical degrees of freedom corresponding to the propagating graviton. The lapse function and shift vectors are perturbed by
\begin{align}
	N=1+\alpha~,\quad N^i=\partial^i\vartheta+\beta^i~,~\text{with}~\partial_i \beta^{i}=0 \,,\label{auxPertDef}
\end{align}
such that they are also irreducible representations of the background $SO(3)$ symmetry. Following the standard protocol of cosmological perturbation theory, we cast \eqref{varphiPertDef}, \eqref{gammaPertDef} and \eqref{auxPertDef} back into the action and vary with respect to the auxiliary fields. This produces constraint equations that we can in principle solve to find the non-dynamical degrees of freedom in terms of the dynamical ones. These solutions can then be inserted back into the action leading to a theory for the dynamical modes only. Such a procedure can produce interactions for the dynamical modes that are not manifest in the original action. For example, if we are interested in the parity-even four-point function of a spectator scalar due to graviton exchange, upon integrating out the non-dynamical parts of the metric one finds a quartic self-interaction for the scalar \cite{Seery:2006vu,Bonifacio:2022vwa}. The same is true for the graviton four-point function where ones finds a quartic self-interaction for the graviton \cite{Bonifacio:2022vwa}. \textit{All} such interactions must be considered to produce a gauge invariant result (or indeed a de Sitter invariant one).

We parametrise the solutions for the non-dynamical modes as
\begin{subequations}
	\begin{align}
\alpha&=C_\alpha(\varphi,\gamma_{ij})=C_\alpha^{(0)} + f^{-1} C_\alpha^{(1)}+ f^{-2} C_\alpha^{(2)}+\cdots~,\\
\vartheta&=C_\vartheta(\varphi,\gamma_{ij})=C_\vartheta^{(0)} + f^{-1} C_\vartheta^{(1)}+ f^{-2} C_\vartheta^{(2)}+\cdots~,\\
		\beta_i&=C_{\beta_i}(\varphi,\gamma_{ij})=C_{\beta_i}^{(0)} + f^{-1} C_{\beta_i}^{(1)}+ f^{-2} C_{\beta_i}^{(2)}+\cdots~ \,,
	\end{align}\label{constraintExpansion}
\end{subequations}
with the zeroth-order solutions in $f^{-1}$ corresponding to those of minimal inflation which were computed in \cite{Maldacena:2002vr,Seery:2006vu,Jarnhus:2007ia}. In principle we need to consider the solutions up to linear order in $f^{-1}$ in order to compute the parity-odd four-point function at that order. Indeed, the solutions at linear order in $f^{-1}$ can be plugged into the zeroth-order part of the action to produce interactions linear in $f^{-1}$, or the zeroth-order solutions can be plugged into the Chern-Simons part of the action to also produce interactions linear in $f^{-1}$. However, as will show in Section \ref{sec4}, we are able to derive our results by relying on the zeroth-order solutions only. The linear solutions and resulting interactions were considered in \cite{GiorgioParity}. 

Plugging the constraints back to the action and eliminating the auxiliary fields, we obtain the gauge-fixed action written purely in terms of the physical degrees of freedom,
\begin{align}
	S=S_\varphi^{(2)}[\varphi]+S_\gamma^{(2)}[\gamma]+S_{\rm int}[\varphi,\gamma]~.
\end{align}
Once we expand the full action around the inflationary solution, the quadratic actions for the inflaton and graviton are given by
\begin{align}
	S_\varphi^{(2)}&=\frac{1}{2}\int \d\eta\d^3 x \, a^2\left[\varphi'^2-(\partial_i \varphi)^2\right]~,\\
	S_\gamma^{(2)}&=\frac{M^2_{\rm{pl}}}{8}\int \d\eta\d^3 x \bigg\{a^{2} \left[\gamma'_{ij}\gamma'_{ij}-\partial_k \gamma_{ij}\partial_k \gamma_{ij}\right]-2 a\kappa\epsilon_{ijk}\left[\gamma'_{il}\partial_j \gamma'_{lk}-\partial_k \gamma_{lj}\partial^2 \gamma_{li}\right]\bigg\}~,\label{linearGravitonAction}
\end{align}
where we have converted to conformal time and $'\equiv\partial_\eta$. All the repeated indices are contractions made with the Kronecker delta. Note that the only explicit time dependence in these two expressions enters via the scale factor. For the Chern-Simons correction this is a consequence of the inflaton shift symmetry for this part of the action. The parameter 
\begin{align}
	\kappa\equiv \frac{\dot{\phi}_0}{f M_{\text{pl}}^2} \,,
\end{align}
characterises the Chern-Simons correction to the graviton quadratic action. Note that such a correction is parity-odd and therefore intrinsically sign-indefinite. In the high-energy limit where $|\partial/a|\gg \kappa^{-1}$, the sign of the kinetic energy term $\mathcal{O}(\gamma'^2)$ in \eqref{linearGravitonAction} becomes negative, leading to a ghost instability problem. Within the classical theory, the ghost instability problem is not a serious issue since its effect on the classical equations of motion is degenerate with a helicity-dependent redefinition of the coupling constant $f\to -f$. However, in a perturbative quantum theory expanded around the linear vacuum, the Hamiltonian is no-longer positive-definite, leading to a potential breakdown of unitarity. In the path integral formulation, this problem manifests itself as the failure of convergence in the path integral that computes the partition function of the quantum theory. The resolution of the ghost instability is, of course, by realising that \eqref{definingAction} is merely a low-energy effective theory valid for energy scales below $|\partial/a|\lesssim\kappa^{-1}$. Any consistent UV completion of Chern-Simons gravity thus requires the presence of other higher-order EFT operators, which could eventually restore the stability of the vacuum in the high-energy limit. In this work, however, we shall adopt an alternative perspective and treat the Chern-Simons term \textit{perturbatively}. In other words, the $\kappa\epsilon \gamma \gamma$ terms in $S_\gamma^{(2)}$ shall be considered as a linear mixing interaction along with the non-linear interactions in $S_{\rm int}$ and accounted for order-by-order. When truncated at any finite order in $\kappa$, the quantum theory is perfectly well-defined and ghost-free.

We are interested in the parity-violating curvature trispectrum sourced by the Chern-Simons term. More specifically, we would like to compute the parity-odd part of the tree-level curvature trispectrum to leading order in the slow-roll parameters $\varepsilon,\eta$ and the inflaton decay constant $f^{-1}$. For simplicity, we shall work in the spatially flat gauge described above and perform a gauge transformation to curvature perturbations at the end. The parity-odd trispectrum must be an odd function of the inflaton decay constant $f$, because gravity is the only source of parity violation in the theory\footnote{We will come back to the case where the Chern-Simons term is not the only source of parity violation later in Section \ref{sec5}.} and a parity transformation is degenerate with a flip $f\to -f$. The leading order contribution must therefore be linear in $f^{-1}$. In addition to the parity-odd linear mixing of the graviton, terms in $S_{\text{int}}$ can also contribute to the parity-odd four-point function at tree-level via $\epsilon \varphi \varphi \gamma$ and $\epsilon \varphi^4$ interactions. Such interactions can arise from the expansion of the Chern-Simons term or after integrating out the non-dynamical modes. This implies that there can be three separate sources to the parity-odd four-point function at this order:
\begin{enumerate}
	\item[I.] The mixing term $\epsilon \gamma \gamma$ in $S_\gamma^{(2)}$, where $f^{-1}\propto \kappa$ comes from a single mixing insertion.
	\item[II.] The interaction vertex $\epsilon \varphi\varphi \gamma$ in $S_{\rm int}$. Here $f^{-1}$ can come from both the vertex directly derived from the Chern-Simons term and the vertex induced by the $C^{(1)}$ constraints in \eqref{constraintExpansion} when plugged into the General Relativity (GR)/inflaton action.\footnote{By GR, we mean the Einstein-Hilbert action of gravity along with its minimal coupling to matter.}
	\item[III.] The (non-local) interaction vertex $\epsilon \varphi\varphi\varphi\varphi$ in $S_{\rm int}$ induced by the constraints. Here $f^{-1}$ comes from the $C^{(1)}$ constraints in \eqref{constraintExpansion} as well as the $C^{(0)}$ constraints when plugged into the original Chern-Simons term in the action.
\end{enumerate}
In the rest of the paper, we will compute each of these contributions. We will see that Contribution I is non-zero and is consistent with our expectation for the shape function from the CCF, while Contributions II and III vanish at the level of the correlator even though they contribute non-trivially to the wavefunction.


\section{Cosmological wavefunctions and correlator-to-correlator factorisation}\label{sec3}

The wavefunction of the universe approach to computing cosmological correlators has gained much attention in recent years since the constraints imposed by key principles such as symmetries, locality and unitarity are more manifest in wavefunction coefficients than in cosmological correlators \cite{Arkani-Hamed:2018kmz,Jazayeri:2021fvk,Goodhew:2020hob}. In this approach one computes the cosmological wavefunctional $\Psi$ that gives the amplitude of finding primordial fields with certain spatial profiles at the end of inflation. For our interests in this work, such a wavefunctional is a functional of $\bar{\varphi}(\bfk) \equiv \varphi(\eta_0, \bfk)$ (the inflaton) and $\bar{\gamma}_{ij}(\bfk) \equiv \gamma_{ij}(\eta_0, \bfk)$ (the graviton). Here $\eta_0 \rightarrow 0$ is the value of conformal time at the end of inflation, c.f. \eqref{quasidSBgsol}. The wavefunctional can be computed via path integral techniques and for weakly coupled theories in perturbation theory, it is dominated by the saddle-point approximation:\footnote{Quantum corrections in the form of loops can be systematically computed in perturbation theory, see e.g. \cite{Cespedes:2023aal} for a nice discussion and \cite{Anninos:2014lwa, Benincasa:2024ptf, Lee:2023jby,Qin:2024gtr} for examples.}
\begin{align}
\Psi[\eta_0; \bar{\varphi}, \bar{\gamma}] \approx e^{i S_{\text{cl}}} \,,
\end{align}
where $S_{\text{cl}}$ is the action of the theory evaluated on the classical solution of interest. For us, this classical solution corresponds to the cosmological solutions of the inflaton and graviton on de Sitter space that satisfy the following boundary conditions:
\begin{align}
\lim_{\eta \rightarrow -\infty} \varphi(\eta, \bfk) &= 0, ~~~ \qquad \lim_{\eta \rightarrow -\infty} \gamma_{ij}(\eta, \bfk) = 0\,, \\
\lim_{\eta \rightarrow \eta_0} \varphi(\eta, \bfk) &= \bar{\varphi}(\bfk), \qquad \lim_{\eta \rightarrow \eta_0} \gamma_{ij}(\eta, \bfk) = \bar{\gamma}_{ij}(\bfk) \,, 
\end{align}
with the vanishing of fields in the far past corresponding to the Bunch-Davies vacuum choice.\footnote{We allow for some evolution in Euclidean time in the far past in order to project onto the vacuum i.e. the far past limit is $\eta \rightarrow - \infty(1 - i \epsilon)$ \cite{Maldacena:2002vr}.} On such classical solutions it is clear that both $S_{\text{cl}}$ and $\Psi$ are functions of the boundary data $\bar{\varphi}$ and $\bar{\gamma}_{ij}$, and in perturbation theory we can express the wavefunctional as an expansion in these spatial profiles.\footnote{Note that we are assuming that the non-dynamical parts of the metric have been integrated out so that the action is a function of the dynamical modes only, as discussed in Section \ref{sec2}.} For our interests in this paper, the expansion of the wavefunctional that is needed to compute the scalar trispectrum at tree-level is 
\begin{align} \label{WavefunctionExpansion}
 \ln \Psi = &  -\frac{1}{2} \int_{\bfk }\psi_{\varphi \varphi}
(k)  \bar{\varphi}(-\bfk) \bar{\varphi}(\bfk) - \frac{1}{2} \sum_{h= \pm} \int_{\bfk} \psi_{\gamma \gamma}^{(h)} (k) \bar{\gamma}_{h}(-\bfk) \bar{\gamma}_{h}(\bfk) \nonumber \\ & + \frac{1}{2} \sum_{h = \pm} \int_{\bfk} \delta \psi_{\gamma \gamma}^{(h)} (k) \bar{\gamma}_{h}(-\bfk) \bar{\gamma}_{h}(\bfk) \nonumber \\ &  + \frac{1}{2} \sum_{h = \pm} \int_{\bfk_1 \bfk_2 \bfk_3}\psi^{(h)}_{\varphi \varphi \gamma} (\{ \bfk \}) \hat{\delta}\left(\sum \bfk \right)  \bar{\varphi}(\bfk_1) \bar{\varphi}(\bfk_2) \bar{\gamma}_{h}(\bfk_3) \nonumber \\ & + \frac{1}{2} \sum_{h = \pm} \int_{\bfk_1 \bfk_2 \bfk_3} \delta \psi^{(h)}_{\varphi \varphi \gamma} (\{ \bfk \}) \hat{\delta}\left(\sum \bfk \right)   \bar{\varphi}(\bfk_1) \bar{\varphi}(\bfk_2) \bar{\gamma}_{h}(\bfk_3) \nonumber \\ & + \frac{1}{24} \int_{\bfk_1 \ldots \bfk_4}  \delta \psi_{\varphi \varphi \varphi \varphi } (\{ \bfk \}) \hat{\delta}\left(\sum \bfk \right)   \bar{\varphi}(\bfk_1) \ldots \bar{\varphi}(\bfk_4) + \ldots \,,
\end{align}
where $(\{ \bfk \})$ indicates a collection of momenta that a wavefunction coefficient depends on, we have converted to the helicity basis using \eqref{GravitonHelicityBasis}, and $+ \ldots$ indicates contributions that we do not need in this work. The graviton quadratic wavefunction coefficients only depend on a single summed helicity since different helicities do not mix at quadratic order. As we discussed in Section \ref{sec2}, the quadratic action of the graviton receives a perturbative correction due to the Chern-Simons term which in turn generates a correction to the graviton quadratic wavefunction coefficient that we have denoted by $\delta \psi_{\gamma \gamma}^{(h)}$. This contribution is linear in the Chern-Simons coupling, $\kappa$. Similarly, we have split up the cubic and quartic wavefunction coefficients into contributions that are zeroth order and linear order in $\kappa$ with any $\delta \psi$ being the linear contribution. Each wavefunction coefficient can be computed using a set of Feynman rules with bulk-boundary and bulk-bulk propagators, and we refer the reader to \cite{Goodhew:2020hob,Benincasa:2022gtd,Benincasa:2022omn,Stefanyszyn:2023qov} if they are not familiar with such rules. 

With the wavefunctional at hand, one can extract cosmological correlators by applying the Born rule. For a mixed correlator with $m$ gravitons and $n-m$ inflatons, we have  
\begin{align} \label{CorrelatorAverage}
\langle \bar{\gamma}_{h_1}(\bfk_{1}) \ldots  \bar{\gamma}_{h_m} (\bfk_{m}) \bar{\varphi}(\bfk_{m+1}) \ldots \bar{\varphi}(\bfk_{n})  \rangle  = \mathcal{N}^{-1} \int \mathcal{D} \bar{\varphi} \mathcal{D} \bar{\gamma} |\Psi|^2  \bar{\gamma}_{h_1}(\bfk_{1}) \ldots \bar{\gamma}_{h_m} (\bfk_{m}) \bar{\varphi}(\bfk_{m+1}) \ldots \bar{\varphi}(\bfk_{n}) \,,
\end{align}
with normalisation 
\begin{align}
\mathcal{N} = \int \mathcal{D} \bar{\varphi} \mathcal{D} \bar{\gamma} |\Psi|^2 \,.
\end{align}
In practice, such an average is computed by expanding the wavefunctional about the dominant Gaussian contributions (the first two terms in \eqref{WavefunctionExpansion}) followed by performing various Gaussian integrals. For the scalar trispectrum at tree-level that we are interested in here i.e. the parity-odd component at linear order in $\kappa$, the resulting expression is
\begin{align} \label{WavefunctionToCorrelator}
\nonumber&B_{\varphi \varphi \varphi \varphi}^{\text{PO}}(\bfk_1,\bfk_2,\bfk_3,\bfk_4) \\
=& \,\frac{1}{ \Pi_{a=1}^{4} \rho_{\varphi \varphi}(k_a)} \left[\delta \rho_{\varphi \varphi \varphi \varphi}(-\bfk_1, -\bfk_2, -\bfk_3, -\bfk_4)   +  \sum_{h = \pm} \frac{\rho^{(h)}_{\varphi \varphi \gamma}(-\bfk_1, -\bfk_2, \mathbf{s}) \,\delta \rho^{(h)}_{\varphi \varphi \gamma}(-\bfk_3, -\bfk_4, -\mathbf{s})}{\rho^{(h)}_{\gamma \gamma}(s)} + 5 ~ \text{perms} \right. \nonumber \\
&\qquad\qquad\qquad\left. 
 +  \sum_{h = \pm} \frac{\rho^{(h)}_{\varphi \varphi \gamma}(-\bfk_1, -\bfk_2, \mathbf{s} ) \,\delta \rho^{(h)}_{\gamma \gamma}(s) \,\rho^{(h)}_{\varphi \varphi \gamma}(-\bfk_3, -\bfk_4, -\mathbf{s})}{\rho^{(h)}_{\gamma \gamma}(s)\,\rho^{(h)}_{\gamma \gamma}(s)} + 2 ~ \text{perms} \right] \,,
\end{align}
where $B$ corresponds to a connected correlator with the delta function stripped off i.e.
\begin{align}
\langle \ldots \rangle = B^{\cdots}(\{ \bfk \}) \hat{\delta} \left( \sum \bfk \right) \,. 
\end{align}
Notice that in \eqref{WavefunctionToCorrelator} it is $\rho(\{ \bfk\}) = \psi(\{ \bfk \}) + \psi^\#(\{ \bfk\})$ that appears (where $\psi^\#(\{ \bfk\})\equiv \psi^{*}(\{ -\bfk \})$ denotes a complex conjugation combined with a parity transformation), since this is the object that appears in $| \Psi |^2$. To see this one needs to use $\bar{\varphi}(\bfk) = \bar{\varphi}^{*}(-\bfk)$ and $\bar{\gamma}(\bfk) = \bar{\gamma}^{*}(-\bfk)$ when computing $\Psi^{*}$. The fact that it is this combination that contributes to the final correlator is the basis of the no-go theorems of \cite{Liu:2019fag,Cabass:2022rhr}. In \eqref{WavefunctionToCorrelator} we have defined $\mathbf{s} = \bfk_{1} + \bfk_{2}$, and the ``$+ ~ 2 ~ \text{perms}$" in the third contribution corresponds to summing over the $s$, $t$ and $u$ channels for $2 \rightarrow 2$ exchange diagrams. The second contribution with ``$+ ~ 5 ~ \text{perms}$"  contains twice as many terms since the two cubic wavefunction coefficients in this term are not the same. Note that the numerical coefficients in front of each channel in \eqref{WavefunctionToCorrelator} are unity thanks to our choice of normalisation in \eqref{WavefunctionExpansion}.    

In Section \ref{sec2} we showed that there are three distinct contributions to the tree-level parity-odd trispectrum in Chern-Simons gravity, with each contribution defined by the vertices that drive the parity violation. We will therefore write
\begin{align}
B_{\varphi \varphi \varphi \varphi}^{\text{PO}} = B_{\varphi \varphi \varphi \varphi}^{\text{PO,I}}+ B_{\varphi \varphi \varphi \varphi}^{\text{PO,II}}+ B_{\varphi \varphi \varphi \varphi}^{\text{PO,III}} \,.
\end{align}
For contribution I, where the parity violation is driven by a parity-odd correction to the graviton two-point function, each of the three structures on the RHS of \eqref{WavefunctionToCorrelator} contribute to the final correlator. Diagrammatically we therefore have\footnote{Notice that in \eqref{ContributionI} we have pulled out an overall minus sign such that the diagrams correspond to $\psi(\{+\bfk\})$. This is valid because we are projecting on the parity-odd component.}
\begin{align} \label{ContributionI}
 B_{\varphi \varphi \varphi \varphi}^{\text{PO,I}} 
    = -\frac{1}{\rho^4_{\varphi\varphi}}&\left[\Re^{\sharp}\raisebox{-17pt}{\includegraphics[scale=0.55]{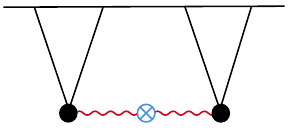}}\,
	+\frac{1}{\rho_{\gamma\gamma}}\,\left(\Re^{\sharp}\raisebox{-17pt}{\includegraphics[scale=0.55]{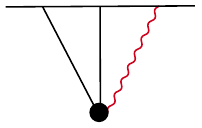}}\times \Re^{\sharp}\raisebox{-17pt}{\includegraphics[scale=0.55]{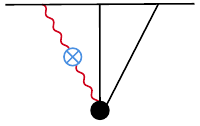}}+\text{5 perms}\right)\right.\nonumber\\
    &\left.~+{\frac{1}{\rho^2_{\gamma\gamma}}}\left(\Re^{\sharp}\raisebox{-17.5pt}{\includegraphics[scale=0.55]{psi3L}}\times \Re^{\sharp}\raisebox{-18pt}{\includegraphics[scale=0.57]{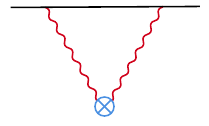}} \times \Re^{\sharp}\raisebox{-17pt}{\includegraphics[scale=0.55]{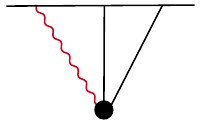}}+\text{2 perms}\right)\right]\,,
\end{align}
where $\Re^{\sharp}$ denotes that we take a wavefunction diagram and add the same diagram but with the result complex conjugated and with all the external momenta flipped sign. In other words, the operation of $\Re^{\sharp}$ takes us from $\psi$ to $\rho$ so diagrams with a $\sharp$ can be considered as diagrams for the computation of $\rho$. In these diagrams the solid lines represent inflaton propagators while wavy lines represent graviton propagators, the solid blobs correspond to parity-even vertices while ``crossed" blobs correspond to parity-odd vertices. Note that the second and third terms take the form of single and double ``propagator cuts" of the first term. For contribution II, where the parity violation is driven by a parity-odd $\varphi \varphi \gamma $ vertex, the last term in the RHS of \eqref{WavefunctionToCorrelator} does not contribute. In this case we therefore have 
\begin{align} \label{WavefunctionII}
 B_{\varphi \varphi \varphi \varphi}^{\text{PO,II}} = -\frac{1}{\rho^4_{\varphi\varphi}}&\left[\Re^{\sharp}\raisebox{-17pt}{\includegraphics[scale=0.55]{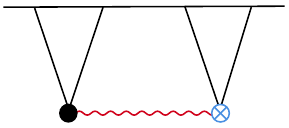}}\,
	+\frac{1}{\rho_{\gamma\gamma}}\,\left(\Re^{\sharp}\raisebox{-17pt}{\includegraphics[scale=0.55]{psi3L}}\times\Re^{\sharp}\raisebox{-17pt}{\includegraphics[scale=0.55]{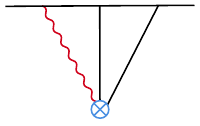}}+\text{5 perms}\right)\right]  \,.
\end{align}
Finally, for contribution III, where the parity violation is driven by a quartic self-interaction of the inflaton, only the first term on the RHS of \eqref{WavefunctionToCorrelator} contributes i.e.
\begin{align} \label{WavefunctionIII}
 B_{\varphi \varphi \varphi \varphi}^{\text{PO,III}} =  -\frac{1}{\rho^4_{\varphi\varphi}}&\left[\Re^{\sharp}\raisebox{-18pt}{\includegraphics[scale=0.57]{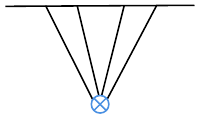}}\right]\,.
\end{align}
In Section \ref{sec4} we will consider each of these three contributions in turn, but let us first summarise our findings.

As we mentioned above, the wavefunction method has some advantages with regards to making some very useful properties of cosmological observables manifest, but for practical computations of cosmological correlators it perhaps seems inefficient since once we have the wavefunction coefficients there is still a functional integral to perform. For us in this work, the wavefunction method is very important since it cleverly packages things together so that individual diagrams have the best possible IR behaviour. This is thanks to the Dirichlet boundary conditions that are placed on bulk-bulk propagators meaning that they vanish at the future conformal boundary thereby softening the IR behaviour of a given diagram. Even if a vertex in the theory appears manifestly IR-divergent i.e. it contains negative powers of conformal time, a corresponding wavefunction exchange diagram could avoid any IR divergences thanks to the vanishing of the bulk-bulk propagator (this is made precise in \cite{Stefanyszyn:2023qov}). We can contrast this with Schwinger-Keldysh diagrammatics \cite{Chen:2017ryl} where internal propagators do not vanish at the future boundary and therefore IR divergences of interaction vertices are not tamed diagram-by-diagram. The wavefunction method essentially takes combinations of Schwinger-Keldysh diagrams such that the sum has softer IR behaviour. Having the ability to diagnose whether a wavefunction diagram is convergent in the IR or not is crucial in the context of computing parity-odd correlators, as pointed out in \cite{Stefanyszyn:2023qov}. In that work we showed that wavefunction coefficients are \textit{helically real} ($\psi^{(h)}(\{ \bfk \}) = \left[\psi^{(-h)}(\{\bfk \} )\right]^*$) if the following assumptions are met:
\begin{itemize}
    \item We work with the tree-level approximation
    \item We work with the Bunch-Davies vacuum choice (fields vanish in the far past)
    \item The external fields are scalars/gravitons and the internal fields are in the complementary or discrete series of de Sitter representations
    \item The time integrals that compute a given wavefunction coefficient are IR-convergent
\end{itemize}
The reality theorem can also be understood as a consequence of unitarity and the CPT symmetry of the de Sitter spacetime, and can be generalised to a phase formula for general external states in $d$-dimensions \cite{Goodhew:2024eup}. Much of our focus in the following sections will be on understanding when this reality theorem holds in Chern-Simons gravity so we can find the non-zero contributions to \eqref{WavefunctionToCorrelator}. We will show, using our reality theorem, that 
\begin{align} \label{VanishingRhos}
\rho_{\varphi \varphi \gamma}^{(h)} \neq 0, \qquad \delta \rho_{\gamma \gamma}^{(h)} \neq 0 , \qquad 
\delta \rho_{\varphi \varphi \varphi \varphi} = 0, \qquad \delta \rho_{\varphi \varphi \gamma}^{(h)} = 0 \,,
\end{align}
meaning that the only contribution to the final parity-odd trispectrum in Chern-Simons gravity is 
\begin{align} \label{WavefunctionToCorrelatorFinal}
B_{\varphi\varphi\varphi\varphi}^{\text{PO}} = \frac{1}{ \Pi_{a=1}^{4} \rho_{\varphi \varphi}(k_a)} \left[ \sum_{h = \pm} \frac{\rho^{(h)}_{\varphi \varphi \gamma}(-\bfk_1, -\bfk_2, \mathbf{s}) \, \delta \rho^{(h)}_{\gamma \gamma}(s)\, \rho^{(h)}_{\varphi \varphi \gamma}(-\bfk_3, -\bfk_4, -\mathbf{s})}{\rho^{(h)}_{\gamma \gamma}(s)\, \rho^{(h)}_{\gamma \gamma}(s)} + 2 ~ \text{perms} \right] \,.
\end{align}
The final trispectrum is therefore \textit{factorised} meaning that there are no total-energy poles. This is just one example of a more general phenomena where parity-odd correlators take factorised forms \cite{Stefanyszyn:2023qov}. As further shown in \cite{Stefanyszyn:2024msm}, if the internal states are in the complementary or discrete series, as is the case here, factorised parity-odd correlators can be written in terms of lower-point correlators i.e. different observables are \textit{linked} in a very non-trivial way. That is indeed the case here for Chern-Simons gravity. To see this we need expressions for $B_{\varphi \varphi \gamma}$, $B_{\gamma \gamma}$ and $\delta B_{\gamma \gamma}$. By performing the relevant Gaussian integrals, we have 
\begin{align}
 B^{(h)}_{\varphi \varphi \gamma} (\bfk_1, \bfk_2, \bfk_3) &= \frac{\rho^{(h)}_{\varphi \varphi \gamma}(-\bfk_1, -\bfk_2, -\bfk_3 )}{\rho_{\varphi \varphi}(k_1)\, \rho_{\varphi \varphi}(k_2)\, \rho^{(h)}_{\gamma \gamma}(k_3)} \,, \label{BphiphigammaFromRhos}\\
 B^{(h)}_{\gamma \gamma}(k) &= \frac{1}{\rho^{(h)}_{\gamma \gamma}(k)} \,, \label{BgammagammaFromRhos}\\ 
 \delta B^{(h)}_{\gamma \gamma}(k) &= \frac{\delta \rho^{(h)}_{\gamma \gamma}(k)}{\rho^{(h)}_{\gamma \gamma}(k) \, \rho^{(h)}_{\gamma \gamma}(k)} \,,
\end{align}
thereby allowing us to write \eqref{WavefunctionToCorrelatorFinal} as
\begin{keyeqn}
\begin{align} \label{CSCCF}
B_{\varphi\varphi\varphi\varphi}^{\text{PO}} (\bfk_1,\bfk_2,\bfk_3,\bfk_4)=  \sum_{h = \pm} \frac{B^{(h)}_{\varphi \varphi \gamma}(\bfk_1, \bfk_2, -\mathbf{s}) \, \delta B^{(h)}_{\gamma \gamma}(s) \, B^{(h)}_{\varphi \varphi \gamma}(\bfk_3, \bfk_4, \mathbf{s})}{B^{(h)}_{\gamma \gamma}(s) \, B^{(h)}_{\gamma \gamma}(s)} + 2 ~ \text{perms} \,.
\end{align}
\end{keyeqn}
This is an example of correlator-to-correlator factorisation (CCF). In \cite{Stefanyszyn:2024msm} we explored examples of such CCF formulae in the context of parity-odd correlators with the parity violation arising due to the mixing between the inflaton and an additional massive spinning field. Our primary aim of this work is to show that such a CCF formula can be realised without the addition of new states during inflation: the inflaton and graviton alone is enough to realise a neat scalar-tensor CCF formula that relates correlators of observational interest. We now turn our attention to the technical details that lead to \eqref{VanishingRhos} and therefore \eqref{CSCCF}. In Section \ref{sec5} we will come back to this CCF formula. 


\section{Explicit calculation of the parity-odd trispectrum}\label{sec4}

In this section, we present the detailed calculation of the parity-odd curvature trispectrum arising from Chern-Simons gravity. We shall work in the spatially flat gauge and compute the inflaton four-point function before gauge-transforming to the curvature trispectrum. As shown in Sections \ref{sec2} and \ref{sec3}, at leading order in slow-roll and the inflaton decay constant $f^{-1}$, there are three possible contributions to the scalar four-point function. We shall consider each of them separately in the following subsections.

\subsection{Contribution I: $\epsilon \gamma\gamma$}\label{contributionISubSect}

The first contribution that we will consider originates from the Chern-Simons correction to the graviton quadratic action. This correction is given by (c.f. \eqref{linearGravitonAction})
\begin{align}
\delta \mathcal{L}_{2}&=-\frac{M^2_{\rm{pl}}\kappa}{4} a(\eta)\, \epsilon_{ijk}\left(\gamma'_{il}\partial_j \gamma'_{lk}-\partial_k \gamma_{lj}\partial^2 \gamma_{li}\right)\,.\label{2ptint}
\end{align}
One might be tempted to add this to the linearised Einstein-Hilbert contribution and solve the resulting equations of motion for arbitrary $\kappa$ therefore yielding the free evolution of the graviton from which we can extract propagators. However, there are at least two reasons why this is not the correct approach. Firstly, the Chern-Simons operator that gives rise to \eqref{2ptint} is only one of the possible higher-derivative operators we can add to the inflationary EFT and to keep control over this tower of operators we must treat corrections to the Einstein-Hilbert action perturbatively. Secondly, the equations of motion arising from the Chern-Simons operator are higher than second order thereby suggesting that the theory has a ghost. This ghost only becomes a problem if the Chern-Simons term is treated non-perturbatively. We will therefore treat \eqref{2ptint} as a perturbative correction to the Einstein-Hilbert action. In practice, this means that we will be interested in computing the parity-odd trispectrum at linear order in $\kappa$ rather than considering the effects of \eqref{2ptint} on the free propagation. In other words, we treat this correction as an ``interaction" in Feynman diagrams. 

In order to produce a four-point function at linear order in $\kappa$, we need an additional interaction and we work with the cubic interaction that is forced to exist by diffeomorphism invariance, namely,
\begin{align}
	\mathcal{L}_{3,\text{int}}&=\frac{1}{2}a^2\gamma_{ij}\partial_i\varphi\partial_j\varphi\,.\label{3ptint}
\end{align}
This is the familiar minimal coupling. With the above interactions, i.e. $\delta \mathcal{L}_{2}$ and $\mathcal{L}_{3,\text{int}}$, Contribution I for the parity-odd trispectrum, in terms of wavefunction diagrams, is given by \eqref{ContributionI}. However, as we will explicitly show below using the reality theorem of \cite{Stefanyszyn:2023qov}, the first two terms in \eqref{ContributionI} vanish such that
\begin{align} 
 B_{\varphi \varphi \varphi \varphi}^{\text{PO,I}} 
   =-{\frac{1}{\rho^4_{\varphi \varphi}}}{\frac{1}{\rho^2_{\gamma\gamma}}}\left(\Re^{\sharp}\raisebox{-17.5pt}{\includegraphics[scale=0.55]{psi3L}}\times \Re^{\sharp}\raisebox{-18pt}{\includegraphics[scale=0.57]{hpsi2}} \times \Re^{\sharp}\raisebox{-17pt}{\includegraphics[scale=0.55]{psi3R}}+\text{2 perms}\right)\,.
\end{align}
Let's start with the first contribution in \eqref{ContributionI}. The corresponding wavefunction coefficient is given by
\begin{align}
	\delta\psi_{\varphi\varphi\varphi\varphi}(\bfk_{1},\bfk_{2},\bfk_3,\bfk_4)
	&=\raisebox{-17pt}{\includegraphics[scale=0.55]{Psi4wMix}}\nonumber\\
	&=-i\,\frac{2 M^2_{\text{pl}}\kappa}{H^5}\sum_{h = \pm} h\,\Big(\bfk_{1}\cdot\hat{\mathbf{e}}^{(h)}(\mathbf{s})\Big)^*\Big(\bfk_{2}\cdot\hat{\mathbf{e}}^{(h)}(\mathbf{s})\Big)^*\Big(\bfk_{3}\cdot\hat{\mathbf{e}}^{(h)}(\mathbf{s})\Big)\Big(\bfk_{4}\cdot\hat{\mathbf{e}}^{(h)}(\mathbf{s})\Big)\nonumber\\
	&~~~\times \,s\int\frac{d\eta_1}{\eta_1^2}\frac{d\eta_2}{\eta_2}\frac{d\eta_3}{\eta_3^2}K_{\varphi}(\eta_1,k_1)K_{\varphi}(\eta_1,k_2)K_{\varphi}(\eta_3,k_3)K_{\varphi}(\eta_3,k_4)\nonumber
	\\
    &~~~\times\left[\partial_{\eta_2}G_\gamma(\eta_1,\eta_2,s)\partial_{\eta_2}G_\gamma(\eta_2,\eta_3,s)-s^2 G_\gamma(\eta_1,\eta_2,s)G_\gamma(\eta_2,\eta_3,s)\right] \nonumber \\ &~~~ + \text{2 perms}\,,\label{psi4mix}
\end{align}
where we have used
\begin{align}
    \epsilon_{ijm}\,k_m{\rm{e}}^{(h)}_{lj}(-\bfk){\rm{e}}^{(h')}_{li}(\bfk)= -2i
    h k\,\delta_{h h'}\,,
\end{align}
here the factor of two comes from the normalisation in the polarisation tensor (\ref{PolarNormal}), so that \eqref{2ptint} simplifies in momentum space to 
\begin{align}
    \mathcal{L}_{2,\text{int}}&= -\frac{M^2_{p}\, \kappa}{2}\sum_{h=\pm} \,a(\eta)\,h k\big(\gamma'^2_h-k^2 \gamma_h^2\big)\,.
\end{align}
In this basis the factor of $h$ breaks the symmetry between the $\pm$ modes and is therefore the source of parity-violation.
The bulk-boundary $K$ and the bulk-bulk $G$ propagators in the wavefunction formalism are given by
\begin{align}
	K_\gamma(\eta,k)&=K_\varphi(\eta,k)=\frac{\varphi^*(k,\eta)}{\varphi^*({k,\eta_0})}=(1-ik\eta)e^{i k\eta}\,,\label{bulktoboundary}\\
	G_\gamma(\eta_1,\eta_2,k)&=P_\gamma(k)\Big[\Big(K^*(\eta_1,k)K(\eta_2,k)\theta(\eta_1-\eta_2)+\left(\eta_1\leftrightarrow\eta_2\right)\Big)-K(\eta_1,k)K(\eta_2,k)\Big]\,.\label{bulktobulk}
\end{align}
In \eqref{psi4mix} we have explicitly written the $s$-channel contribution and $+ \text{2 perms}$ contains the $t$ and $u$ channel contributions. Since we treat the Chern-Simons correction perturbatively, the graviton propagators are insensitive to helicity and are the standard de Sitter ones. Notably, the last term in (\ref{bulktobulk}) ensures that the bulk-bulk propagator vanishes at the conformal boundary where $\eta_0 \rightarrow 0$. This will play a crucial role in our analysis. In this sense, the wavefunction approach corrects the usual Feynman propagator that arises in the in-in and Schwinger/Keldysh formalisms thereby cleverly arranging the perturbation theory into the diagrams that have the best possible IR behaviour. As we mentioned in Section \ref{sec3}, wavefunction coefficients are \textit{purely real} at tree level under the assumptions of scale invariance, Bunch-Davies vacuum conditions and IR convergence of the time integrals \cite{Stefanyszyn:2023qov}. We are indeed working with scale invariant interactions and have imposed Bunch-Davies vacuum conditions to arrive at the propagators in \eqref{bulktoboundary} and \eqref{bulktobulk},\footnote{The Bunch-Davies condition imposes that the propagators goes like $e^{i k \eta}$ in the far past.} so in order to conclude that \eqref{psi4mix} is a purely real function of the external kinematics, without having to actually compute the nested time integrals, we need to check if the time integrals are IR-convergent. This is a very non-trivial question given the negative powers of $\eta_1$, $\eta_2$ and $\eta_3$ in \eqref{psi4mix}. Indeed, the vertices that contribute to this diagram are \textit{not} manifestly IR-convergent meaning that the interaction vertices themselves diverge at the boundary:
\begin{align}
	S_{3, \text{int}} \sim \int \d \eta\,a^2(\eta)\varphi(\eta) \varphi(\eta) \gamma (\eta) &\sim \int \d \eta\, \eta^{-2}\,, \\
   \delta S_{2} \sim  \int \d \eta\,a(\eta)\gamma (\eta)\gamma (\eta) &\sim \int\d \eta\, \eta^{-1}\,.
\end{align}
However, this by itself does not imply that \eqref{psi4mix} diverges at the boundary because there is other time dependence to consider coming from the propagators and as we have already discussed above, the bulk-bulk propagator decays as we approach the boundary. To see this more clearly, we can rewrite the bulk-bulk propagator (\ref{bulktobulk}) as \cite{Cabass:2022rhr} 
\begin{align}
	G_\gamma(\eta_1,\eta_2,k)=-2iP_\gamma(k) K_\gamma(\eta_2,k)\Im K_\gamma(\eta_1,k)\theta(\eta_1-\eta_2)+(\eta_1\leftrightarrow \eta_2)~.\label{bulktobulk2}
\end{align}
Consider the case where $\eta_1 \rightarrow 0$ meaning that the only contribution comes from the $\theta(\eta_1 - \eta_2)$ term. We then have
\begin{align}
	\lim_{\eta_1\rightarrow 0}G_\gamma(\eta_1,\eta_2,k)\sim\Im K_\gamma (\eta_1)\sim\eta_1^3\,.\label{bulkIR} 
\end{align}
Thanks to this nice property, (\ref{psi4mix}) can indeed be IR-convergent even if the vertices are not manifestly so. Now that we see that convergence is at least in principle possible, we will check the actual IR scaling of \eqref{psi4mix}. Given the different theta functions in the integrand, there are $3!=6$ possible orderings. However, the integrand is symmetric in $\eta_1$ and $\eta_3$ so we can assume $\eta_1>\eta_3$ without loss of generality and consider the resulting $3$ time orderings.

\begin{figure}
	\centering
	\includegraphics[width=1\linewidth]{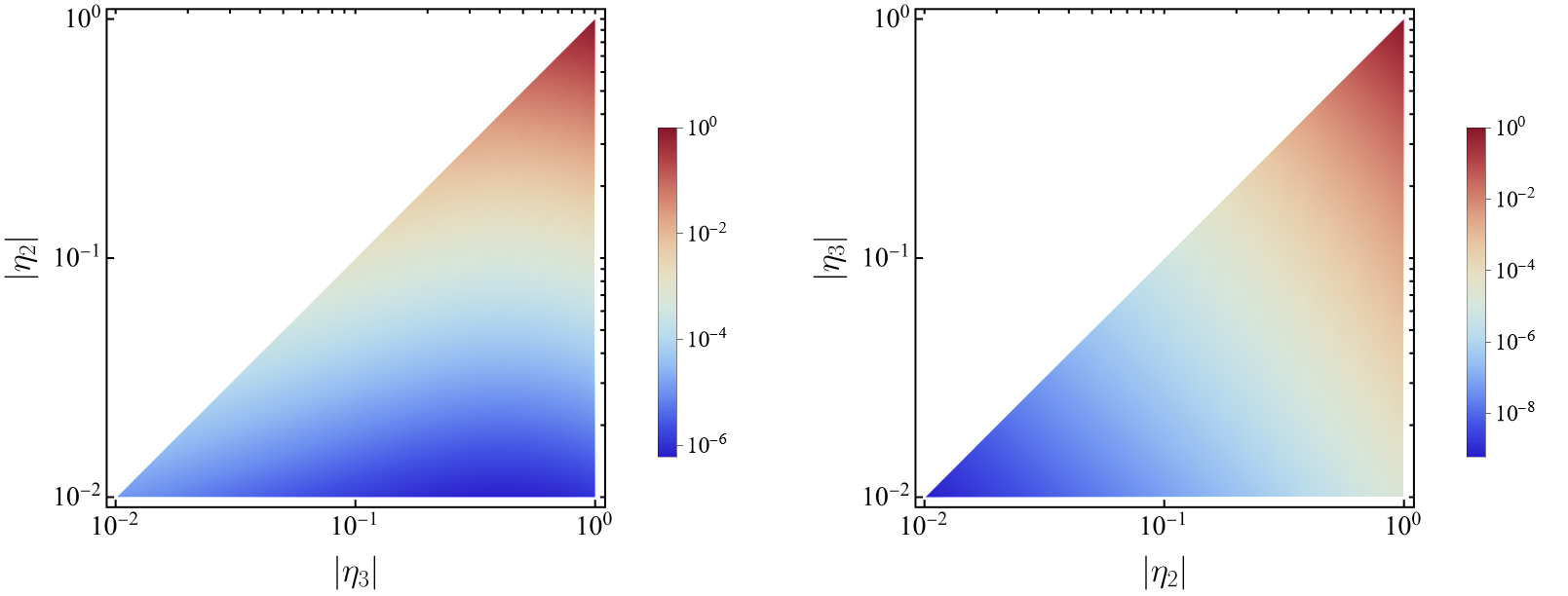}
	\caption{The absolute value of the integrand of (\ref{psi4mix}). For better visualisation, $\eta_1$ has been integrated out, and the density plot shows its dependence on the two remaining vertex times. Here we have set $k_1=k_3=2$, $k_2=k_4=\sqrt{s^2+k_1^2}$ and $s=1$. The \textit{left} and \textit{right} panels correspond to the time orderings $\eta_2>\eta_3$ and $\eta_3>\eta_2$ respectively.}
	\label{integrandIR}
\end{figure}

\paragraph{IR scaling of integrands}
To examine the IR behaviour of the integrand, the following late-time behaviours are useful
\begin{align}
K(\eta)\,\sim\,\eta^0,\qquad K'(\eta)\,\sim\,\eta\,, \qquad \Im K\,\sim \eta^3\,, \qquad\Im K'\,\sim\,\eta^2\,.\label{IRimK}
\end{align}
Note that there is no term linear in $\eta$ in the expansion of $K$ which is why $K'(\eta)\,\sim\, \eta$ rather than $K'(\eta)\,\sim\, \eta^0$ as naively expected. This is the basis of the Manifest Locality Test of \cite{Jazayeri:2021fvk}.
\begin{enumerate}
	\item[$\bullet$] $\eta_1>\eta_2>\eta_3$. We begin with this time ordering which comes from the $\theta (\eta_1 -\eta_2) \theta(\eta_2 - \eta_3)$ part of the integrand. Since $\eta_3$ is the earliest time, we consider its integration first up to $\eta_2$:\footnote{Note that when the time derivative acts on the bulk-bulk propagator, the delta functions that arise from the derivative acting on the theta functions cancel out such that the derivative only acts on the mode functions (for higher order derivatives this would not be the case and there would indeed be delta functions).} 
    \begin{align} 
  &  \int \frac{d \eta_1}{\eta_1^2} \frac{d \eta_2}{\eta_2} \int^{\eta_2}\frac{d \eta_3}{\eta^2_3}K^2_{\eta_1}K^2_{\eta_3} [\partial_{\eta_2}  K_{\eta_2} \Im K_{\eta_1} \partial_{\eta_2} \Im K_{\eta_2} K_{\eta_3}-s^2  K_{\eta_2} \Im K_{\eta_1}  \Im K_{\eta_2} K_{\eta_3}] \nonumber \\
    \sim & \int \frac{d \eta_1}{\eta_1^2} \frac{d \eta_2}{\eta_2} K^2_{\eta_1}[\partial_{\eta_2}  K_{\eta_2} \Im K_{\eta_1} \partial_{\eta_2} \Im K_{\eta_2} -s^2  K_{\eta_2} \Im K_{\eta_1}  \Im K_{\eta_2} ] \eta_2^{-1} \,.
    \end{align}
  We then integrate $\eta_2$ up to $\eta_1$ followed by integrating $\eta_1$ to the boundary to find
      \begin{align} 
  & \int \frac{d \eta_1}{\eta_1^2} \int^{\eta_1} \frac{d \eta_2}{\eta_2} K^2_{\eta_1}[\partial_{\eta_2}  K_{\eta_2} \Im K_{\eta_1} \partial_{\eta_2} \Im K_{\eta_2} -s^2  K_{\eta_2} \Im K_{\eta_1}  \Im K_{\eta_2} ] \eta_2^{-1} \nonumber \\
  & \sim \int^{{\eta_0}} d \eta_1 K^2_{\eta_1}[ \Im K_{\eta_1} -s^2 \Im K_{\eta_1} ] \sim \mathcal{O}(\eta_0^4) \,.
    \end{align} 
We therefore see that this part of the full integral is IR-convergent. Here we have dropped the momentum-dependence of the propagators since they are not important for this discussion and have indicated the time dependence via subscripts to keep the expressions compact. 
    
\item[$\bullet$] $\eta_1>\eta_3>\eta_2$. We repeat this procedure for this time ordering which comes from the $\theta(\eta_1 - \eta_2) \theta(\eta_3-\eta_2)$ part of the integrand. We again find $\mathcal{O}(\eta_0^4)$ behaviour so again we have IR-convergence. 

\item[$\bullet$] $\eta_2>\eta_1>\eta_3$. For this final case the relevant part of the integrand has $\theta(\eta_2 - \eta_1) \theta(\eta_2 - \eta_3)$. We again find covergence but this time with a $\mathcal{O}(\eta_0^2)$ behaviour in the IR. 
\end{enumerate}
All three time orderings are therefore IR-convergent which ensures that the full wavefunction coefficient in \eqref{psi4mix} is IR-convergent. To complement this analytical argument for convergence, we also confirm the safe IR behaviour of $\delta \psi _{\varphi\varphi\varphi\varphi}$ numerically and present the result Figure \ref{integrandIR}. We first finish a layer of integration on the left vertex $\eta_1$, and then generate a contour plot of the integrand with the remaining two time variables $\eta_2$ and $\eta_3$. The results clearly illustrate that there is no IR divergence.  

Now that we have verified IR convergence of the time integrals that compute $\delta \psi_{\varphi \varphi \varphi \varphi}$, we can impose the reality theorem of \cite{Stefanyszyn:2023qov} which states that $\Im \delta \psi_{\varphi\varphi\varphi\varphi}=0$. Recall that it is actually $\delta \rho_{\varphi \varphi \varphi \varphi}$ that contributes to the correlator (c.f. \eqref{WavefunctionToCorrelator}) and given that we are working with the parity-odd sector we have
\begin{align}
	\delta\rho_{\varphi\varphi\varphi\varphi}(\bfk_{1},\bfk_{2},\bfk_{3},\bfk_{4})&\equiv \delta\psi_{\varphi\varphi\varphi\varphi}(\bfk_{1},\bfk_{2},\bfk_{3},\bfk_{4})+\delta\psi^*_{\varphi\varphi\varphi\varphi}(-\bfk_{1},-\bfk_{2},-\bfk_{3},-\bfk_{4})\nonumber\\
	&= \delta\psi_{\varphi\varphi\varphi\varphi}(\bfk_{1},\bfk_{2},\bfk_{3},\bfk_{4})-\delta\psi^*_{\varphi\varphi\varphi\varphi}(\bfk_{1},\bfk_{2},\bfk_{3},\bfk_{4})\nonumber\\
	&=0\,,
\end{align}
where in the second line, we used the fact that $\delta \psi _{\varphi\varphi\varphi\varphi}$ is parity-odd i.e. $\delta\psi_{\varphi\varphi\varphi\varphi}(\{-\bfk\})=-\delta\psi_{\varphi\varphi\varphi\varphi}(\{\bfk\})$, and in the third line we used that $\Im \delta \psi_{\varphi\varphi\varphi\varphi}=0$. We note that $\delta\psi_{\varphi\varphi\varphi\varphi}(\{-\bfk\})=-\delta\psi_{\varphi\varphi\varphi\varphi}(\{\bfk\})$ is not manifest in \eqref{psi4mix} but becomes manifest once we use the relation
\begin{align}
&\sum_{h = \pm} h\,\Big(\bfk_{1}\cdot\hat{\mathbf{e}}^{(h)}(\mathbf{s})\Big)^*\Big(\bfk_{2}\cdot\hat{\mathbf{e}}^{(h)}(\mathbf{s})\Big)^*\Big(\bfk_{3}\cdot\hat{\mathbf{e}}^{(h)}(\mathbf{s})\Big)\Big(\bfk_{4}\cdot\hat{\mathbf{e}}^{(h)}(\mathbf{s})\Big)\nonumber\\
&= i\,\hat{\bf{s}}\cdot(\bfk_2\times\bfk_4)\Big[\bfk_1\cdot\bfk_3-(\bfk_{1}\cdot\hat{\bf{s}})(\bfk_{3}\cdot\hat{\bf{s}})\Big]\,.\label{PolarSum}
\end{align}
We therefore see that the contribution \eqref{psi4mix} to the wavefunction drops out once we compute the correlator. In terms of the Schwinger/Keldysh diagrams, this cancellation corresponds to cancellations between ``black" and ``white" diagrams. For readers who are not familiar with the results of \cite{Stefanyszyn:2023qov}, we will now review the main arguments that lead to the reality of $\delta \psi_{\varphi\varphi\varphi\varphi}$.

\paragraph{Reality of $\delta\psi_{\varphi\varphi\varphi\varphi}$}
Having shown that the time integrals that compute \eqref{psi4mix} are convergent, we can Wick-rotate all time variables clockwise by $90^{\circ}$ in the complex plane such that we integrate along the positive imaginary axis rather than the negative real axis.\footnote{The result of the integration does not change by Cauchy's theorem: we close the contour and drop the arc at infinity thanks to the Bunch-Davies vacuum condition. See \cite{Cabass:2022rhr,Stefanyszyn:2023qov} for details.} Such a rotation enables us to easily verify the parts of the \eqref{psi4mix} that are real/imaginary. We will denote the rotated time variables as $\chi$, defined by
\begin{align}
	\eta_i=i\,e^{i\epsilon}\chi_i\,,\qquad \epsilon \rightarrow 0^{+}~\text{and}~\chi>0\,.
\end{align}
We can then analyse each part of \eqref{psi4mix}:
\begin{enumerate}
	\item[$\bullet$] Polarisation sum. The polarisation factor, with the sum over helicity, satisfies
	\begin{align}
	&	\sum_{h = \pm} h\,\Big(\bfk_{1}\cdot\hat{\mathbf{e}}^{(h)}(\mathbf{s})\Big)^*\Big(\bfk_{2}\cdot\hat{\mathbf{e}}^{(h)}(\mathbf{s})\Big)^*\Big(\bfk_{3}\cdot\hat{\mathbf{e}}^{(h)}(\mathbf{s})\Big)\Big(\bfk_{4}\cdot\hat{\mathbf{e}}^{(h)}(\mathbf{s})\Big) \nonumber \\
      =- & \left[\sum_{h = \pm} h\,\Big(\bfk_{1}\cdot\hat{\mathbf{e}}^{(h)}(\mathbf{s})\Big)^*\Big(\bfk_{2}\cdot\hat{\mathbf{e}}^{(h)}(\mathbf{s})\Big)^*\Big(\bfk_{3}\cdot\hat{\mathbf{e}}^{(h)}(\mathbf{s})\Big)\Big(\bfk_{4}\cdot\hat{\mathbf{e}}^{(h)}(\mathbf{s})\Big)  \right]^{*}\,,
	\end{align}
    which follows from \eqref{UnitVector} (i.e. a helicity flip is the same as complex conjugation) so the polarisation factor is purely imaginary.

\item[$\bullet$] Bulk-boundary propagator. This propagator is a purely real function of the rotated variable:
	\begin{align}
		K(\chi,k)=(1+k\,\chi) e^{-k\,\chi}\,\in\,\mathbb{R}\,.
	\end{align}
	\item[$\bullet$] Bulk-bulk propagator. As with the bulk-boundary propagators, the bulk-bulk propagator is a function of $i \eta_1$ and $i \eta_2$ and therefore it is also a real function after the Wick rotations.
  \item[$\bullet$] Integration measures. Each integration measure reads $i \int \frac{d \eta_i}{\eta_i^4}$ with the factor of $\eta_{i}^{-4}$ coming from $\sqrt{-g} = a^{4}(\eta)$ and with the factor of $i$ coming from the fact we are computing $e^{iS}$. The integration measures are therefore purely real after the rotations.    
  \item[$\bullet$] Derivatives. Time derivatives enter as $\eta_i \partial_{\eta_i}$ with the factor of $\eta_i$ dictated by scale invariance. In momentum space, spatial derivatives yield factors of $i \eta_i \bfk$ again with the factor of $\eta_i$ forced by scale invariance. Both time and spatial derivatives thefore yield purely real factors after Wick rotation.   
\end{enumerate}
We therefore see that all factors in \eqref{psi4mix} are purely real after Wick rotation other than the polarisation factor which is purely imaginary. This combines with the overall factor of $i$ thereby ensuring that the full $\delta \psi_{\varphi \varphi \varphi \varphi}$ is a purely real function of the external kinematics, as promised. 

Let's now turn our attention to the second term in \eqref{ContributionI} which consists of a product of a contact wavefunction diagram and a single-exchange one. To make our analysis more transparent, we define an \textit{indexed} wavefunction coefficient by extracting any polarisation tensors. For example, for the single-exchange diagram, we have
\begin{align}
	\delta \psi^{(h)}_{\varphi\varphi\gamma}(\mathbf{s},\bfk_{3},\bfk_{4}) &= \raisebox{-17.5pt}{\includegraphics[scale=0.55]{Psi3RwMix}}\nonumber\\
	&\equiv \mathrm{e}^{(h)}_{ij}(\mathbf{s})\,\delta \psi^{ij}_{\varphi\varphi\gamma}(\mathbf{s},\bfk_{3},\bfk_{4})\,.
\end{align}
This indexed wavefunction coefficient for this exchange diagram is computed by 
\begin{align}
    \delta\psi^{ij}_{\varphi\varphi\gamma}=& -\,\frac{\sqrt{2}M^2_{\rm{pl}}\kappa s}{H^3} \sum_{h = \pm} h\,\left(\mathrm{e}^{(h)}_{ij}(\mathbf{s})\right)^*\Big(\bfk_{3}\cdot\hat{\mathbf{e}}^{(h)}(\mathbf{s})\Big)\Big(\bfk_{4}\cdot\hat{\mathbf{e}}^{(h)}(\mathbf{s})\Big)\int\frac{d\eta_2}{\eta_2} \frac{d\eta_3}{\eta_3^2}K_\varphi(\eta_3,k_3)K_\varphi(\eta_3,k_4)\nonumber\\
    &\times \Big[\partial_{\eta_2}K_\gamma(\eta_2,s)\partial_{\eta_2} G_{\gamma}(\eta_2,\eta_3,s)- s^2 K_\gamma(\eta_2,s)G_\gamma(\eta_2,\eta_3,s)\Big]\,.\label{psi3ij}
\end{align}
We again want to apply the reality theorem to this index wavefunction coefficient and to do so we need to check if the integrals are IR convergent. The vertices are not manifestly so, as before, but the bulk-bulk propagator can in principle save the day.
\paragraph{IR scaling of integrands}
Since the integrand (\ref{psi3ij}) has two vertices, there are only two possible time orderings coming from the $\theta(\eta_2-\eta_3)$ and $\theta(\eta_3-\eta_2)$ factors in the bulk-bulk propagator. Note that $\eta_2$ is the vertex associated with the quadratic correction to the graviton action while $\eta_3$ is associated with the minimal coupling vertex.
\begin{enumerate}
    \item[$\bullet$] $\eta_2>\eta_3$. First, let's consider the $\theta(\eta_2-\eta_3)$ part and first integrate $\eta_3$ to $\eta_2$ followed by integrating $\eta_2$ to $\eta_0$:
    \begin{align}
      & \int \frac{d \eta_2}{\eta_2} \int^{\eta_2} \frac{d \eta_3}{\eta_3^2} K_{\eta_3} K_{\eta_3}[\partial_{\eta_2}K_{\eta_2} \partial_{\eta_2} \Im K_{\eta_2} K_{\eta_3} - s^2 K_{\eta_2} \Im K_{\eta_2} K_{\eta_3}] \nonumber \\
       \sim & \int^{\eta_0} \frac{d \eta_2}{\eta^2_2} [\partial_{\eta_2}K_{\eta_2} \partial_{\eta_2} \Im K_{\eta_2}  - s^2 K_{\eta_2} \Im K_{\eta_2} ] \sim \mathcal{O}(\eta_0^2) \,. 
    \end{align}
 This contribution is therefore convergent. 
    \item[$\bullet$] $\eta_2<\eta_3$. We can repeat the analysis for this time ordering coming from the $\theta(\eta_3-\eta_2)$ part of the integrand. In this case we find $\mathcal{O}(\eta_0^2)$ behaviour so again we have convergence. 
\end{enumerate}
This full wavefunction coefficient is therefore convergent (we have verified this fact numerically, as above) and since we have exact scale invariance and are working with Bunch-Davies vacuum conditions, we can apply the reality theorem of \cite{Stefanyszyn:2023qov} to conclude that
\begin{align}
	 \delta\tilde{\psi}^{ij}_{\varphi\varphi\gamma}(\mathbf{s},\bfk_{3},\bfk_{4})=\left[\delta\tilde{\psi}^{ij}_{\varphi\varphi\gamma}(\mathbf{s},\bfk_{3},\bfk_{4})\right]^*\,.
\end{align}
This can be seen directly from \eqref{psi3ij} given our discussion above. After Wick rotating the two time variables, we find that the time integral yields a purely imaginary result since we have an odd number of $\eta_2$ factors. However, the polarisation factor is also purely imaginary thanks to the factor of $h$ in the sum. The product of the two, and therefore the full wavefunction coefficient, is then a real function of the kinematics.  
Again from \eqref{WavefunctionToCorrelator} it is $\delta \rho^{(h)}_{\varphi \varphi \gamma}$ that contributes to the correlator and we have 
\begin{align}
	\delta\rho^{(h)}_{\varphi\varphi\gamma}(\mathbf{s},\bfk_{3},\bfk_{4}) &=	\delta\psi^{(h)}_{\varphi\varphi\gamma}(\mathbf{s},\bfk_{3},\bfk_{4})+\delta\psi^{*(h)}_{\varphi\varphi\gamma}(-\mathbf{s},-\bfk_{3},-\bfk_{4})\nonumber\\
	&= \mathrm{e}^{(h)}_{ij}(\mathbf{s})\left[\delta\psi^{ij}_{\varphi\varphi\gamma}(\mathbf{s},\bfk_{3},\bfk_{4})-\delta\psi^{ij*}_{\varphi\varphi\gamma}(\mathbf{s},\bfk_{3},\bfk_{4})\right]\nonumber\\
	& = 0\,,
\end{align}
where we used \eqref{PolarReality} to factor out the polarisation tensor and we picked up a relative minus sign since we are working with the parity-odd sector. We therefore conclude that the second term in \eqref{psi4mix} vanishes. 

As we mentioned at the start of this section, we are therefore left with only one of the three terms in \eqref{psi4mix}. We will now compute this term and show that it is indeed non-zero. There are two different wavefunction coefficients that are relevant here. There is the parity-even wavefunction coefficient arising from the minimal coupling of the inflaton to gravity, which was first computed by Maldacena in \cite{Maldacena:2002vr}, which we have denoted by $\psi_{\varphi \varphi \gamma}^{(h)}$, and the parity-odd correction to the graviton two-point function $\delta \psi_{\gamma \gamma}^{(h)}$. Both diagrams are contact ones and are therefore straightforward to compute. We start with the three-point function that is computed by
\begin{align}
	\psi^{(h)}_{\varphi\varphi\gamma}(\bfk_{1},\bfk_{2},-\mathbf{s}) &= \raisebox{-17.5pt}{\includegraphics[scale=0.55]{psi3L}}\nonumber\\
	&=-i\sqrt{2}\Big(\bfk_{1}\cdot\hat{\mathbf{e}}^{(h)}(\mathbf{s})\Big)^*\Big(\bfk_{2}\cdot\hat{\mathbf{e}}^{(h)}(\mathbf{s})\Big)^*\int_{-\infty}^{\eta_0} \frac{d\eta}{H^2 \eta^2} K_{\varphi}(\eta,k_1)\,K_{\varphi}(\eta,k_2)\, K_{\gamma}(\eta,s)\nonumber\\
	&=-i\sqrt{2}\Big(\bfk_{1}\cdot\hat{\mathbf{e}}^{(h)}(\mathbf{s})\Big)^*\Big(\bfk_{2}\cdot\hat{\mathbf{e}}^{(h)}(\mathbf{s})\Big)^*(1-k_1\partial_{k_1})(1-k_2\partial_{k_2})(1-s\partial_{s})\int_{-\infty}^{\eta_0} \frac{d\eta}{H^2 \eta^2} e^{i E_L\eta}\nonumber\\
	&= \frac{\sqrt{2}}{H^2}\Big(\bfk_{1}\cdot\hat{\mathbf{e}}^{(h)}(\mathbf{s})\Big)^*\Big(\bfk_{2}\cdot\hat{\mathbf{e}}^{(h)}(\mathbf{s})\Big)^*\times \left[\frac{1}{E_L^2}\left(-E_L^3+E_L\,e_2+e_3\right)+\frac{i}{\eta_0}\right]\,,\label{psi3phiphigamma}
\end{align}
where we have introduced the following kinematic variables
\begin{align}
	E_L= k_1+k_2+s\,,\qquad e_2=k_1 k_2 +k_1 s+k_2 s\,, \qquad e_3=k_1 k_2 s\,.
\end{align}
The last term with $\eta_0^{-1}$  in (\ref{psi3phiphigamma}) shows that the wavefunction coefficient $\psi^{(h)}_{\varphi\varphi\gamma}$ in GR is IR-divergent \cite{Maldacena:2002vr}. Here we do not have the bulk-bulk propagator to provide any IR softening so the fact that the interaction is not manifestly convergent does indeed lead to a divergent wavefunction. This divergence resides in the phase of the wavefunction and therefore drops out when we convert to the correlator. Indeed,
\begin{align}
	\rho_{\varphi\varphi \gamma}^{(h)}(\bfk_{1},\bfk_{2},-\mathbf{s})&=\psi^{(h)}_{\varphi\varphi\gamma}(\bfk_{1},\bfk_{2},-\mathbf{s})+\psi^{*(h)}_{\varphi\varphi\gamma}(-\bfk_{1},-\bfk_{2},-\mathbf{s})\nonumber\\
	&=\frac{2\sqrt{2}}{H^2}\Big(\bfk_{1}\cdot\hat{\mathbf{e}}^{(h)}(\mathbf{s})\Big)^*\Big(\bfk_{2}\cdot\hat{\mathbf{e}}^{(h)}(\mathbf{s})\Big)^*\times \left[\frac{1}{E_L^2}\left(-E_L^3+E_L\,e_2+e_3\right)\right]\,.
\end{align}
By relabelling the external momenta, we also have 
\begin{align}
	\rho_{\varphi\varphi \gamma}^{(h)}(\bfk_{3},\bfk_{4},\mathbf{s})
	=\frac{2\sqrt{2}}{H^2}\Big(\bfk_{3}\cdot\hat{\mathbf{e}}^{(h)}(\mathbf{s})\Big)\Big(\bfk_{4}\cdot\hat{\mathbf{e}}^{(h)}(\mathbf{s})\Big)\times\left[\frac{1}{E_R^2}\left(-E_R^3+E_R\,\tilde{e}_2+\tilde{e}_3\right)\right]\,,
\end{align}
with 
\begin{align}
	E_R= k_3+k_4+s\,,\qquad \tilde{e}_2=k_3 k_4 +k_3 s+k_4 s\,, \qquad \tilde{e}_3=k_3 k_4 s\,.
\end{align}
The only thing left for us to compute is the correction to the graviton two-point function and we find
\begin{align}
	\delta\psi^{(h)}_{\gamma\gamma} &= \raisebox{-17.5pt}{\includegraphics[scale=0.55]{hpsi2}}\nonumber\\
	&=i\frac{M^2_{\rm{pl}}\kappa\,s\,h}{ H} \int_{-\infty}^{\eta_0} \frac{d\eta}{\eta}\Big[K'_\gamma(\eta,s)K'_\gamma(\eta,s)-s^2\,K_\gamma(\eta,s)K_\gamma(\eta,s)\Big]\nonumber\\
	&=\frac{M^2_{\rm{pl}}\kappa\,s^3\,h}{ H}\left[\frac{\pi}{2}+i\left(\frac{3}{2}-\gamma_{\rm{E}}-\log(-2s\eta_0)\right)\right]\,.\label{deltaPsigammagammah}
\end{align}
Again, we see that this wavefunction coefficient is IR-divergent\footnote{We note that here the IR divergence appears at two-point level as a correction to $\psi_{\gamma\gamma}$ because the treatment for the mixing vertex is strictly perturbative. In fact, it can be shown that by including higher-orders of mixing insertions, such logarithmic divergences can be resummed into a convergent correction to the two-point wavefunction coefficient $\psi_{\gamma\gamma}$. In this case it is the divergence in the three-point wavefunction coefficient $\psi_{\varphi \varphi \gamma}$ that ultimately allows us to bypass the no-go theorems and find a non-zero result for the parity-odd trispectrum.} and it also yields a non-zero $\delta \rho_{\gamma \gamma}^{(h)}$:
\begin{align}
	\delta\rho^{(h)}_{\gamma\gamma}=\frac{h M^2_{\rm{pl}}\pi\kappa s^3}{ H}\,.
\end{align}
If we compute the corresponding correction to the graviton power spectrum then we indeed see that the divergence in the wavefunction does not make it to the correlator which is consistent with the findings of \cite{Creminelli:2014wna}.

We can now put everything together to compute the third term in \eqref{psi4mix} and therefore the full contribution I. Using
\begin{align}
	\left(\bfk_{1}\cdot \mathbf{\mathrm{e}}^{(h)}(\mathbf{s})\right)^*\left(\bfk_{3}\cdot\mathbf{\mathrm{e}}^{(h)}(\mathbf{s})\right)=\frac{1}{2}\Big[\bfk_{1}\cdot\bfk_{3}-(\bfk_{1}\cdot\hat{\mathbf{s}})(\bfk_{3}\cdot\hat{\mathbf{s}})+i h \hat{\mathbf{s}}\cdot (\bfk_{1}\times\bfk_{3}) \Big]\,,
\end{align}  
we find
\begin{keyeqn}
	\begin{align} 
		B_{\varphi \varphi \varphi \varphi}^{\text{PO,I}}  
		=&-i\,\frac{\pi\kappa H^7}{2M^2_{\rm{pl}}}\frac{[\hat{\bfk}_1\cdot\hat{\bfk}_3-(\hat{\bfk}_{1}\cdot\hat{\bf{s}})(\hat{\bfk}_{3}\cdot\hat{\bf{s}})]}{(k_1 k_2 k_3 k_4)^2}\frac{\hat{\bf{s}}\cdot(\hat{\bfk}_2\times\hat{\bfk}_4)}{s^3 E^2_L E^2_R} \left(E_L^3-E_L\,e_2-e_3\right) \left(E_R^3-E_R\,\tilde{e}_2-\tilde{e}_3\right)\nonumber\\
		&+\text{2 perms.} \label{POFinalTrispectrum}
	\end{align} 
\end{keyeqn}
One can easily check that the full result is purely imaginary, as it should be, scales as $k^{-9}$ as dictated by scale invariance, and has no total-energy singularities as expected from the results of \cite{Stefanyszyn:2023qov}. As we mentioned in the introduction, the result we get differs from that found in \cite{Creque-Sarbinowski:2023wmb}.

\subsection{Contribution II:  $\epsilon \varphi\varphi\gamma$ and III: $\epsilon\varphi\varphi\varphi\varphi$}

Turning to Contributions II and III, we will show in this subsection that both of them vanish at leading order in slow-roll. The proof essentially makes use of the parity-violation no-go theorems in \cite{Liu:2019fag,Cabass:2022rhr,Thavanesan:2025kyc}, and involves checking that the no-go theorems do indeed apply here, in particular, checking IR convergence of the interactions (as we did above for Contribution I). We shall proceed in three steps where we first briefly review the no-go theorems. Then we establish a lemma on the on-shell classical action and pin down the origin of the interactions of interest from the constraint solutions. Finally, we use general scaling arguments to show that the interactions are IR-convergent and therefore drop out of the parity-odd trispectrum (we are able to do this without explicitly computing the precise form of the interactions).

\subsubsection*{No-go theorems on parity violation}

The no-go theorems state that in a theory of massless fields with de Sitter mode functions, assuming unitarity, locality, Bunch-Davies vacuum, scale invariance and the manifest IR convergence of interactions,\footnote{By manifest IR convergence, we mean that the interaction operator behaves in the late-time limit $\eta\to 0$ as $\int \d\eta \,\eta^\alpha$ with $\Re \alpha>-1$, so that its contribution to the action is always finite regardless of the boundary condition of the fields.} the parity-odd $n$-point correlators vanish at tree level \cite{Cabass:2022rhr,Liu:2019fag}. Since we are treating the Chern-Simons term perturbatively, the free-theory mode functions are massless and de Sitter-invariant. Unitarity and locality are retained as long as we treat the Chern-Simons term perturbatively and truncate it at a finite order. In single-field slow-roll inflation, scale invariance is softly broken by the weak time dependence in the background. However, when expanded to leading order in slow roll parameters, scale invariance of the Lagrangian is exact. We therefore only need to verify the manifest IR-convergence of the interactions in order to apply the no-go theorems to Chern Simons gravity.

\subsubsection*{A lemma on the on-shell wavefunction} \label{lemmaSection}

To check manifest IR convergence, we need to know the form and origin of the $\epsilon \varphi\varphi\gamma$ and $\epsilon\varphi\varphi\varphi\varphi$ vertices. These interactions can come from plugging in the $C^{(0)}$ constraint solutions into the Chern Simons action expanded around the inflationary solution, or from plugging the $C^{(1)}$ constraint solutions into the GR part of the action (see \eqref{constraintExpansion} for definitions of $C^{(0)}$ and $C^{(1)}$). We now show that the latter is impossible at linear order, i.e. $\mathcal{O}(f^{-1})$.

As outlined in Section \ref{sec3}, in the wavefunction formalism, one performs a path integral over all possible histories of cosmic evolution to obtain the wavefunction of the universe given by
\begin{align}
    \Psi[\bar\Phi]=\int_{\rm BD}^{\Phi(\eta_0)=\bar\Phi} \mathcal{D}\Phi \,e^{i S[\Phi]}\,,
\end{align}
where $\Phi=\varphi,\gamma_{ij},\cdots$ collectively denotes all the fields active during inflation. We integrate over fields that satisfy Bunch-Davies vacuum conditions in the far past, meaning that the classical solutions vanish there, and take on some spatial profile in the far future. In the semi-classical limit, the path integral is dominated by a classical saddle which leaves the action stationary, i.e.
\begin{align}
    \Psi[\bar\Phi]= e^{i S[\Phi_{\rm cl}(\bar{\Phi})]}\left[1+\mathcal{O}(\hbar)\right]\,,
\end{align}
where $\Phi_{\rm cl}$ satisfies the full equation of motion,
\begin{align}
	\left.\frac{\delta S}{\delta \Phi}\right|_{\Phi=\Phi_{\rm cl}}=0~, \quad \Phi_{\rm cl}(\eta_0)=\bar{\Phi}~.
\end{align}
All tree-level processes are therefore nicely packaged into the on-shell classical action $S[\Phi_{\rm cl}]$, and even at this semi-classical level, computations at higher orders in perturbation theory can become very involved. However, if we are only interested in the result at first order in a coupling constant, the computation can be simplified. Consider a theory consisting of two parts,
\begin{align}
    S=S_1+\lambda S_2~,
\end{align}
where we have explicitly pulled out a small coupling constant $|\lambda|\ll 1$. Now there are two routes we can take to compute the wavefunction:
\begin{enumerate}
    \item \textit{Route A}: The first way is to compute the classical solution $\Phi_{\rm cl}$ to the full theory subjected to the boundary conditions,
\begin{align}
    \left.\frac{\delta(S_1+\lambda S_2)}{\delta \Phi}\right|_{\Phi=\Phi_{\rm cl}}=0~,\label{FullEoM}
\end{align}
and obtain the classical action
\begin{align}
    S_1^{\rm cl}&\equiv S_1[\Phi_{\rm cl}]~,\\
    S_2^{\rm cl}&\equiv S_2[\Phi_{\rm cl}]~.
\end{align}
The semi-classical wavefunction is then given by $e^{i (S_1^{\rm cl}+\lambda S_2^{\rm cl})}$, which can then be putatively expanded in powers of $\lambda S_2[\Phi_{\rm cl}]$,
\begin{align}
    \Psi_{\rm cl}=e^{i (S_1^{\rm cl}+\lambda S_2^{\rm cl})}=e^{i S_1^{\rm cl}}+i \lambda\, e^{i S_1^{\rm cl}} S_2^{\rm cl}-\frac{\lambda^2}{2} e^{i S_1^{\rm cl}} (S_2^{\rm cl})^2+\mathcal{O}(\lambda^3)~.
\end{align}
Notice that this is \textit{not} a uniform expansion in powers of $\lambda$, since the classical solution $\Phi_{\rm cl}$ contains $\lambda$ to all orders. We stress that this Route A is, by definition, the \textit{canonical} way to compute the semi-classical wavefunction.
    \item \textit{Route B}: The second way is to expand the integrand of the path integral first,
    \begin{align}
        \Psi=\int\mathcal{D}\Phi \,e^{i (S_1+\lambda S_2)}=\int\mathcal{D}\Phi \,e^{i S_1}+i \lambda\,\int\mathcal{D}\Phi \, e^{i S_1} S_2-\frac{\lambda^2}{2} \int\mathcal{D}\Phi \,e^{i S_1} S_2^2+\mathcal{O}(\lambda^3)~,
    \end{align}
    and then perform the saddle-point approximation for each of these terms separately,
    \begin{align}
        \hat{\Psi}_{\rm cl}=e^{i \hat{S}_1^{\rm cl}}+i \lambda\, e^{i \hat{S}_1^{\rm cl}} \hat{S}_2^{\rm cl}-\frac{\lambda^2}{2} e^{i \hat{S}_1^{\rm cl}} (\hat{S}_2^{\rm cl})^2+\mathcal{O}(\lambda^3)~,\label{RouteBPsihatExpansion}
    \end{align}
    where the classical actions
    \begin{align}
        \hat{S}_1^{\rm cl}&\equiv S_1[\hat{\Phi}_{\rm cl}]~,\\
        \hat{S}_2^{\rm cl}&\equiv S_2[\hat{\Phi}_{\rm cl}]~,
    \end{align}
    are obtained from a classical solution $\hat{\Phi}_{\rm cl}$ that extremises the action $S_1$ alone,
    \begin{align}
    \left.\frac{\delta S_1}{\delta \Phi}\right|_{\Phi=\hat{\Phi}_{\rm cl}}=0 \,.\label{PartialEoM}
    \end{align}
    Note that the expansion \eqref{RouteBPsihatExpansion} is indeed a uniform expansion in powers of $\lambda$. The advantage of Route B is that the computation is technically much simpler due to a simpler equation of motion \eqref{PartialEoM}. The disadvantage is that it is expected to be less accurate than Route A since the separate saddle-point approximations in \eqref{RouteBPsihatExpansion} is not well-justified.\footnote{In principle, the saddles must be determined from $\delta(S_1 +\ln S_2)/\delta\Phi=0$ etc.. However, these equations are ill-defined and the only way to make sense of them is to resum them back to Route A. As a compromise, one picks up a wrong saddle ($\hat{\Phi}_{\rm cl}$) which is nevertheless not far from the correct one ($\Phi_{\rm cl}$).}
        
\end{enumerate}
We now claim that $\hat{\Psi}_{\rm cl}$ agrees with $\Psi_{\rm cl}$ at linear order in $\lambda$, but differs at quadratic order. To show this, we take their difference,
\begin{align}
    \Psi_{\rm cl}-\hat{\Psi}_{\rm cl}=\Big(e^{i S_1^{\rm cl}}-e^{i \hat{S}_1^{\rm cl}}\Big)+i \lambda \Big(e^{i S_1^{\rm cl}} S_2^{\rm cl}-e^{i \hat{S}_1^{\rm cl}} \hat{S}_2^{\rm cl}\Big)+\mathcal{O}(\lambda^2)~.\label{ABDiff}
\end{align}
Comparing \eqref{FullEoM} with \eqref{PartialEoM}, we deduce that the classical solutions differ by
\begin{align}
    \Phi_{\rm cl}-\hat{\Phi}_{\rm cl}=\mathcal{O}(\lambda)~.
\end{align}
This leads to a first-order difference in the classical action for $S_2$,
\begin{align}
    S_2^{\rm cl}-\hat{S}_2^{\rm cl}=S_2[\hat{\Phi}_{\rm cl}+\mathcal{O}(\lambda)]-S_2[\hat{\Phi}_{\rm cl}]=\mathcal{O}(\lambda)~.
\end{align}
The difference in $S_1$, however, starts at second order thanks to the variational principle \eqref{PartialEoM},
\begin{align}
    S_1^{\rm cl}-\hat{S}_1^{\rm cl}=S_1[\hat{\Phi}_{\rm cl}+\mathcal{O}(\lambda)]-S_1[\hat{\Phi}_{\rm cl}]=\left.\frac{\delta S_1}{\delta\Phi}\right|_{\Phi=\hat{\Phi}_{\rm cl}}\times \mathcal{O}(\lambda)+\mathcal{O}(\lambda^2)=\mathcal{O}(\lambda^2)~.
\end{align}
This means that both terms in \eqref{ABDiff} are quadratic in $\lambda$, meaning that $\hat{\Psi}_{\rm cl}$ matches $\Psi_{\rm cl}$ at first order in $\lambda$, i.e.
\begin{align}
	    \Psi_{\rm cl}-\hat{\Psi}_{\rm cl}=\mathcal{O}(\lambda^2)~.
\end{align}
It is important to note that the error starts non-zero at second order, since $\Phi_{\rm cl}$ contains bulk-bulk propagators joining two $S_2$ vertices whereas $\hat{\Phi}_{\rm cl}$ does not. In other words, causality between the two $S_2$ vertices is washed out in Route B. We also note that the agreement of $\Psi_{\rm cl}$ and $\hat\Psi_{\rm cl}$ at $\mathcal{O}(\lambda)$ is a reminiscence of first-order time-independent perturbation theory in quantum mechanics, where the shift in the energy level is computed at first order via the zeroth-order energy eigenstates. To summarise, we establish the following lemma:
\begin{keythrm}
	\begin{lemma}
		{\rm\textbf{(First-order approximation)}} The semi-classical wavefunction $\Psi_{\rm cl}$ corresponding to a theory $S=S_1+\lambda S_2$ is well-approximated to first-order in $\lambda$ by evaluating the full action at the partial saddle defined by $\delta S_1=0$.\label{kTRealityTheorem}
	\end{lemma}
\end{keythrm}

Returning to the discussion of the Chern-Simons gravity model, we set $\lambda=f^{-1}$ and $S_2=-1/4\int \phi R\widetilde{R}$. Then this lemma implies that at first order in $f^{-1}$ for the on-shell wavefunction, the graviton stays on the shell of the GR/inflaton action. Henceforth, we are allowed to use the GR constraints for the auxiliary fields and drop the higher-order terms (c.f. \eqref{constraintExpansion}),
\begin{subequations}
	\begin{align}
		\alpha&\approx C_\alpha^{(0)}(\varphi,\gamma_{ij}) ~,\\
		\vartheta&\approx C_\vartheta^{(0)}(\varphi,\gamma_{ij}) ~,\\
		\beta_i&\approx C_{\beta_i}^{(0)}(\varphi,\gamma_{ij}) ~.
	\end{align}\label{constraintApproximation}
\end{subequations}
The theory of perturbations at $\mathcal{O}(f^{-1})$ therefore solely comes from plugging these solutions into the Chern Simons part of the action. If we wanted to compute the wavefunction at higher orders in $f^{-1}$ then we would indeed need to solve the constraints including the corrections due to the Chern Simons action but this is not necessary for our interests in this work.

\subsubsection*{Manifest IR convergence}

We are now in a position to check the IR convergence of the interaction vertices involved in Contributions II and III. Due to exact scale invariance at leading order in slow roll, we can simply extract the IR behaviour of an interaction vertex from power counting techniques. More specifically, we will be interested in the power of conformal time $\eta$ in the interaction vertices in the late-time limit. For instance, for a vertex that goes as
\begin{align}
	S_{\rm int}\supset \int \d\eta \,\eta^\alpha\times (\cdots)~,\quad \eta\to 0~,
\end{align}
manifest IR convergence is assured if $\Re \alpha>-1$. The IR scaling of some familiar quantities are given by
\begin{align}
	a(\eta)\sim \eta^{-1}~,\quad \varphi,\gamma_{ij}\sim \eta^0~,\quad \dot\varphi,\dot \gamma_{ij}\sim \eta^2~.
\end{align}
For the time derivatives, one factor of $\eta$ comes from the inverse scale factor when converting to conformal time and the second comes from taking the derivative of the mode functions with respect to conformal time. In position space, the inflaton and the graviton transform under dilations as\footnote{Such equations should be understood as only being valid when sandwiched between de Sitter-invariant states, such as the Bunch-Davies vacuum. Here the graviton does not carry a conformal dimension because we have pulled out the scale factors in \eqref{gammaPertDef}.}
\begin{align}
	\varphi(\lambda \eta,\lambda \mathbf{x})=\varphi(\eta,\mathbf{x})~,\quad \gamma_{ij}(\lambda \eta,\lambda \mathbf{x})=\gamma_{ij}(\eta,\mathbf{x})\,.
\end{align}
Following \cite{Seery:2006vu}, we expand the auxiliary fields using the GR/inflation constraints $C^{(0)}$ in powers of perturbations,
\begin{subequations}
	\begin{align}
		\alpha = \sum_{m = 1}^\infty \alpha_m ~, \quad
		\vartheta = \sum_{m = 1}^\infty \vartheta_m ~, \quad
		\beta_j = \sum_{m = 1}^\infty \beta_{mj} ~.
	\end{align}
\end{subequations}
The linear solutions are given by \cite{Maldacena:2002vr} 
\begin{align}
	\alpha_1 = \frac{\dot{\phi}_0}{2H}  \varphi
	\quad \text{and} \quad
	\beta_{1j} = 0 ~.\label{alpha1Sol}
\end{align}
We notice that $\alpha_1$ scales as a scalar field under dilation with a IR scaling equivalent to that of a massless field,
\begin{align}
	\alpha_1(\lambda \eta,\lambda \mathbf{x})=\alpha_1(\eta,\mathbf{x})~,\quad \alpha_1\sim \eta^0~.
\end{align}
The linear solution to $\vartheta$ satisfies
\begin{align}
	\frac{4H}{a^2} \partial^2 \vartheta_1 &=
	-2 V' \varphi
	-2\dot{\phi}_0 \dot{\varphi} +
	2\alpha_1(-6H^2 + \dot{\phi}_0^2)\nonumber\\
	& =(-2V'-6H \dot{\phi}_0)\varphi+\frac{\dot{\phi}_0^3}{H}\varphi-2\dot{\phi}_0 \dot{\varphi}\nonumber\\
	& = -2\dot{\phi}_0 \dot{\varphi}+\cdots~,
\end{align}
where in the second line, we have applied the linear solution \eqref{alpha1Sol} to $\alpha$, and in the third line, we have applied the background slow-roll equations. Terms higher order in slow-roll parameters have been neglected in the ellipsis. Thus to leading-order in slow-roll, we have
\begin{align}
	\vartheta_1=-\frac{\dot{\phi}_0}{2H}a^2\partial^{-2} \dot\varphi~,
\end{align}
and $\vartheta_1$ also transforms as a massless scalar,
\begin{align}
	\vartheta_1(\lambda \eta,\lambda \mathbf{x})=\vartheta_1(\eta,\mathbf{x})~,\quad \vartheta_1\sim \eta^0~.\label{vartheta1Scaling}
\end{align}
We also note that terms higher-order in slow roll could behave differently and become IR-divergent in the late-time limit, thereby contributing non-trivially to the trispectrum. However, they are nevertheless suppressed by slow-roll parameters as long as the \textit{final} trispectrum starts out non-zero at lowest order, which we have already shown to be the case in Section \ref{contributionISubSect}. 

Moving on to quadratic terms in the constraints, we have \cite{Jarnhus:2007ia,Seery:2006vu}
\begin{align}
	\alpha_2 &= \frac{\alpha_1^2}{2} + \frac{1}{2H} \partial^{-2}
	\left(  \partial_j \dot{\varphi} \partial_j \varphi
	  + \dot{\varphi} \partial^2 \varphi  + \frac{1}{a^2} \partial^2 \alpha_1 \partial^2 \vartheta_1 -
	\frac{1}{a^2} \partial_i \partial_j \alpha_1 \partial_i \partial_j
	\vartheta_1 \right)\\
	&=\frac{1}{2H} \partial^{-2}
		\left(  \partial_j \dot{\varphi} \partial_j \varphi
		  + \dot{\varphi} \partial^2 \varphi\right)~,
\end{align}
where we have again dropped terms high-order in slow-roll (i.e. $\mathcal{O}(\dot{\phi}_0^2)$ using \eqref{alpha1Sol}). The remaining terms are not slow-roll suppressed since they correspond to the minimal coupling of a scalar field quanta to gravity. Therefore its scaling behaviour reads
\begin{align}
	\alpha_2(\lambda \eta,\lambda \mathbf{x})=\alpha_2(\eta,\mathbf{x})~,\quad \alpha_2\sim \eta^2~.\label{alpha2Scaling}
\end{align} 
Likewise, the leading quadratic terms in $\beta_{j}$ are not slow-roll suppressed,
\begin{align}
	\beta_{2j} \simeq 2a^2\partial^{-4} \left(
	\partial_m \partial_j \dot{\varphi}   \partial_m \varphi   +
	\partial_j \dot{\varphi}   \partial^2 \varphi  
	- \partial^2 \dot{\varphi}   \partial_j \varphi   -
	\partial_m \dot{\varphi}   \partial_m \partial_j \varphi
	\right)~,
\end{align}
leading to the scaling behaviour
\begin{align}
	\beta_{2j}(\lambda \eta,\lambda \mathbf{x})=\lambda^{-1}\beta_{2j}(\eta,\mathbf{x})~,\quad \beta_{2j}\sim \eta^0~.
\end{align}
The quadratic terms in $\vartheta$ are
\begin{align}
	\vartheta_2 =
	\frac{1}{4H}\partial^{-2}\Bigg[ &- \left(a^2\dot{\varphi}^2  + \partial_i \varphi  \partial_i \varphi\right)  + \frac{1}{a^2} \left(\partial^2 \vartheta_1 \partial^2 \vartheta_1 -\partial_i \partial_j \vartheta_1 \partial_i \partial_j
	\vartheta_1\right)  +
	2 \dot{\phi}_0 \partial_i \vartheta_1 \partial_i
	\varphi  \nonumber\\ & + 4 \alpha_1 \left(2H \partial^2 \vartheta_1 +
		a^2 \dot{\phi}_0  \dot{\varphi}  \right)
	 -
	a^2(3 \alpha_1^2 - 2 \alpha_2)(-6H^2 + \dot{\phi}_0^2 )-
		a^2 V'' \varphi^2\Bigg]~.
\end{align}
With \eqref{vartheta1Scaling}, it is straightforward to check that terms in the first line scale at most as $\eta^0$ and the first two are not slow-roll suppressed. The second line, on the other hand, appears to be IR-divergent. However, we note that in the slow-roll phase of inflation, both $\alpha_1^2\sim \mathcal{O}(\dot{\phi}_0^2)$ and $V''\sim \mathcal{O}(\dot{\phi}_0^2,\ddot{\phi}_0)$ are higher-order in slow-roll parameters. The remaining term proportional to $\alpha_2$ is IR-convergent due to \eqref{alpha2Scaling}. As a result, we have
\begin{align}
	\vartheta_2(\lambda \eta,\lambda \mathbf{x})=\vartheta_2(\eta,\mathbf{x})~,\quad \vartheta_2\sim \eta^0~.\label{vartheta2Scaling}
\end{align}
In summary, all the constrained fields have the expected conformal dimension under dilation and are IR finite.

After examining the scaling behaviours of the GR/inflation constraints, we can now proceed with the analysis of Contribution II and Contribution III without performing explicit calculations. The reasoning shall be based on the following four observations:
\begin{enumerate}
	\item[(i).] Since we are working at leading order in slow roll, we only need to keep one factor of $\dot{\phi}_0$. Indeed, in Section \ref{contributionISubSect} we have already found a non-zero contribution at this order. 
	
	\item[(ii).] The full Weyl tensor $W_{\mu\nu\rho\sigma}$ contains two derivatives. However, when expanded upon the de Sitter background \eqref{quasidSBgsol}, the time derivatives may hit the time-dependent background, leading to fewer derivatives acting on perturbations. At linear order in perturbations, this only happens in the vector and tensor sectors. It is straightforward to check that the scalar sector contains at least two spatial derivatives,
	\begin{align}
		W\supset \partial^2 \alpha,~\partial^2 \vartheta,~\partial^2 \dot\vartheta~,
	\end{align}
	the vector sector contains at least one spatial derivative,
	\begin{align}
			W\supset \partial^2\beta,~\partial \dot\beta,~\partial \beta~,
		\end{align}
	and the tensor sector contains either two general derivatives or a single time derivative,
	\begin{align}
			W\supset \partial^2 \gamma,~\partial\dot\gamma,~ \ddot{\gamma},~\dot{\gamma}~.
	\end{align}

	\item[(iii).] At quadratic order in perturbations in the Lagrangian, fields with different helicities do not mix into each other due to the rotational symmetry.
	
	\item[(iv).] There is no parity-violating scalar cubic interaction in the Lagrangian due to kinematics, i.e. three momenta are always coplanar and blind to parity.
\end{enumerate}

Let us begin with analysing Contribution II, where we need two inflaton fields and a graviton field. Expanding the inflaton field into background and perturbations, we expect two sub-contributions,
\begin{itemize}
	\item $\dot{\phi}_0 t \,( W \cdot\widetilde W)$.
	\vskip 1pt
	For this sub-contribution, the two inflatons and a graviton have to come from the two Weyl tensors. Given that the Weyl tensor vanishes on the background, this boils down to four classes: $(\mathcal{S}_1 \cdot \mathcal{S}_1 \gamma)$, $(\mathcal{S}_1 \mathcal{S}_1 \cdot \gamma )$ and $(\mathcal{S}_2\cdot\gamma)$, $(\beta_2\cdot\gamma)$, where $\mathcal{S}_n=\alpha_n,\vartheta_n$ are the $n$-th order scalar constraints. Since there is already an overall $\dot{\phi}_0$ and $\mathcal{S}_1\propto \dot{\phi}_0$, the first two classes are slow-roll suppressed and can be discarded. The last two classes are impossible due to the absence of helicity-mixing stated in (iii).
	
	\item $\varphi \,(W \cdot \widetilde W)$
	\vskip 1pt
	For this sub-contribution, there is already an inflaton outside, hence the Weyl tensors need to provide another inflaton and a graviton. The only possibility is $\varphi\, (\mathcal{S}_1\cdot \gamma)$. As mentioned in (ii)., the first Weyl tensor brings a scalar constraint $\mathcal{S}_1$ with two spatial derivatives. To balance the spatial indices in the presence of a Levi-Civita symbol, we need at least one spatial derivative from the other Weyl tensor. The only contribution is therefore $W\supset \partial\dot\gamma$. The resultant interaction operators are either
	\begin{align}
		S_{\rm int}\supset \int\d \eta \d^3 x\, a \epsilon\varphi (\partial^2 \mathcal{S}_1 \cdot\partial\dot\gamma)\sim \int\d \eta \,\eta^1
	\end{align}
	or
	\begin{align}
		S_{\rm int}\supset \int\d \eta \d^3 x\, a \epsilon\varphi (\partial^2 \dot{\vartheta_1} \cdot\partial\dot\gamma)\sim \int\d \eta \,\eta^3~,
	\end{align}
	which are perfectly IR-convergent.
\end{itemize}
Turning to Contribution III, we again expect two sub-contributions,
\begin{itemize}
	\item $\dot{\phi}_0 t \,( W \cdot\widetilde W)$.
	\vskip 1pt
	For this sub-contribution, we need four inflatons from the two Weyl tensors. Due to the outstanding $\dot{\phi}_0$, the cases $(\mathcal{S}_1\cdot \widetilde{W})$ and $(\mathcal{S}_1 \mathcal{S}_1\cdot \widetilde{W})$ are slow-roll suppressed and can be discarded. The remaining possibilities are $(\mathcal{S}_2\cdot \mathcal{S}_2)$ and $(\beta_2\cdot \beta_2)$. The scalar-scalar case $(\mathcal{S}_2\cdot \mathcal{S}_2)$ is impossible due to the absence of any parity-violation in the scalar sector at this order. The vector-vector case $(\beta_2\cdot \beta_2)$, on the other hand, can exist but with an odd number of spatial derivatives to balance the spatial indices. Since the vector sector of Weyl tensor contains at least one spatial derivative, the structure of the interactions must be either
	\begin{align}
		S_{\rm int}\supset \int\d \eta \d^3 x\, a^{-1}\epsilon\partial\beta_{2}\partial^2\beta_{2}\sim \int\d \eta \,\eta^1
	\end{align}
	or
	\begin{align}
			S_{\rm int}\supset \int\d \eta \d^3 x\, a^{-1}\epsilon\partial\dot{\beta}_{2}\partial^2\beta_{2}\sim \int\d \eta \,\eta^2~,
	\end{align}
	both of which are IR-convergent.	
	
	\item $\varphi \,(W \cdot \widetilde W)$
	\vskip 1pt
	For this sub-contribution, there is already an inflaton outside, hence one of the Weyl tensors must provide a single inflaton while the other must provide two. There are three possibilities: $\varphi\, (\mathcal{S}_1\cdot \mathcal{S}_1 \mathcal{S}_1)$, $\varphi\, (\mathcal{S}_1\cdot \mathcal{S}_2)$ or $\varphi\, (\mathcal{S}_1\cdot \beta_2)$. The first possibility is excluded due to too much slow-roll suppression, whereas the second possibility is excluded by (iv). The last possibility is in fact admissible. The Levi-Civita symbol now compels us to have an even number of spatial derivatives. Again thanks to the structure of the Weyl tensor, the interaction must be of the form
	\begin{align}
		S_{\rm int}\supset \int\d \eta \d^3 x\, a^{-1}\epsilon\partial^2 \mathcal{S}_1\partial^2\beta_{2}\sim \int\d \eta \,\eta^1~,
	\end{align}
	or
	\begin{align}
		S_{\rm int}\supset \int\d \eta \d^3 x\, a^{-1}\epsilon\partial^2 \dot{\vartheta}_1\partial^2\beta_{2}\sim \int\d \eta \,\eta^3~,
	\end{align}
	again both IR-convergent.
\end{itemize}
Consequently, all interactions involved in Contribution II and III are IR-convergent at leading order in slow-roll and therefore we can indeed apply the no-go theorems of \cite{Liu:2019fag,Cabass:2022rhr} to Contributions II and III in Chern Simons gravity.

\subsubsection*{Summary}

To summarise, by the no-go theorem on parity violation, both Contribution II and III exactly vanish at leading order in slow-roll,
\begin{keyeqn}
    \begin{align}
    B_{\varphi\varphi\varphi\varphi}^{\rm PO,II}=B_{\varphi\varphi\varphi\varphi}^{\rm PO,III}=0~.
\end{align}
\end{keyeqn}
The same result was arrived at in \cite{GiorgioParity} by explicitly finding the interactions and performing the in-in computation of the resulting trispectrum. 
\subsection{Going to the $\zeta$-gauge}

So far we have computed everything in the spatially flat gauge with the inflaton fluctuation $\varphi$ taking the lead for the scalar perturbations. What is more relevant for cosmological observations is the statistics of curvature perturbations. To translate the inflaton correlators into curvature correlators, we apply a gauge transformation
\begin{align}
	\nonumber\bar{\varphi}(\mathbf{k})=-\frac{\dot{\phi}_0}{H}\,\Bigg[\bar\zeta(\mathbf{k})&+ \frac{1}{2}\int_{\mathbf{k}=\mathbf{q}_1+\mathbf{q}_2} A(\mathbf{k},\mathbf{q}_1,\mathbf{q}_2)\bar{\zeta}(-\mathbf{q}_1)\bar{\zeta}(-\mathbf{q}_2)\\
	&+ \frac{1}{6}\int_{\mathbf{k}=\mathbf{q}_1+\mathbf{q}_2+\mathbf{q}_3} B(\mathbf{k},\mathbf{q}_1,\mathbf{q}_2,\mathbf{q}_3)\bar{\zeta}(-\mathbf{q}_1)\bar{\zeta}(-\mathbf{q}_2)\bar{\zeta}(-\mathbf{q}_3)+\cdots\Bigg]~.
\end{align}
Here $A$ and $B$ are coefficient functions that follow from performing a diffeomorphism to the constant-inflaton surface. To leading order in derivatives, the diffeomorphism only involves a reparametrisation of time as measured by the background clock $\phi_0(t)$ \cite{Jarnhus:2007ia,Arroja:2008ga}, $A$ and $B$ are insensitive to the Chern-Simons term and are parity-even by definition. This also agrees with the intuition from the separate universe approach \cite{Wands:2000dp}, since spatial gradients are suppressed by inflationary expansion.\footnote{In principle, one cannot exclude the possibility of having enough inverse Laplacians to overcome the spatial gradients, which can only possibly originate from the Chern-Simons correction for the constrained fields. However such terms are forbidden by the lemma in Section \ref{lemmaSection}. In addition, their kinematic singularities would be distinct due to the absence of time integrals.} The classical saddle of the wavefunctional of the universe now becomes
\begin{align} \label{WavefunctionExpansionZeta}
 \ln \Psi = &  -\frac{1}{2} \int_{\bfk }\psi_{\zeta \zeta}
(k)  \bar{\zeta}(-\bfk) \bar{\zeta}(\bfk) - \frac{1}{2} \sum_{h= \pm} \int_{\bfk} \psi_{\gamma \gamma}^{(h)} (k) \bar{\gamma}_{h}(-\bfk) \bar{\gamma}_{h}(\bfk) \nonumber \\ & + \frac{1}{2} \sum_{h = \pm} \int_{\bfk} \delta \psi_{\gamma \gamma}^{(h)} (k) \bar{\gamma}_{h}(-\bfk) \bar{\gamma}_{h}(\bfk) \nonumber \\ &  + \frac{1}{2} \sum_{h = \pm} \int_{\bfk_1 \bfk_2 \bfk_3}\psi^{(h)}_{\zeta \zeta \gamma} (\{ \bfk \}) \hat{\delta}\left(\sum \bfk \right)  \bar{\zeta}(\bfk_1) \bar{\zeta}(\bfk_2) \bar{\gamma}_{h}(\bfk_3) \nonumber \\ & + \frac{1}{2} \sum_{h = \pm} \int_{\bfk_1 \bfk_2 \bfk_3} \delta \psi^{(h)}_{\zeta \zeta \gamma} (\{ \bfk \}) \hat{\delta}\left(\sum \bfk \right)   \bar{\zeta}(\bfk_1) \bar{\zeta}(\bfk_2) \bar{\gamma}_{h}(\bfk_3) \nonumber \\ & + \frac{1}{24} \int_{\bfk_1 \ldots \bfk_4}  \delta \psi_{\zeta \zeta \zeta \zeta } (\{ \bfk \}) \hat{\delta}\left(\sum \bfk \right)   \bar{\zeta}(\bfk_1) \ldots \bar{\zeta}(\bfk_4) + \ldots \,,
\end{align}
which is now a function of $\zeta$ and $\gamma$. The relevant wavefunction coefficients schematically change by
\begin{subequations}
	\begin{align}
		\psi_{\zeta \zeta}&=\left(\frac{\dot{\phi}_0}{H}\right)^2\psi_{\varphi \varphi}~,\\
		\psi_{\zeta \zeta \gamma}&=\left(\frac{\dot{\phi}_0}{H}\right)^2\psi_{\varphi \varphi \gamma}~,\\
		\delta\psi_{\zeta \zeta \gamma}&=\left(\frac{\dot{\phi}_0}{H}\right)^2 \delta\psi_{\varphi \varphi \gamma}~,\\
		\delta\psi_{\zeta \zeta \zeta \zeta}&=\left(\frac{\dot{\phi}_0}{H}\right)^4 \delta\psi_{\varphi \varphi \varphi \varphi}- 6\left(\frac{\dot{\phi}_0}{H}\right)^3 A\cdot\psi_{\varphi \varphi \varphi}+ \left(\frac{\dot{\phi}_0}{H}\right)^2 \left(4 B+3A\cdot A\right)\cdot\psi_{\varphi \varphi}~,
	\end{align}
\end{subequations}
where $(\cdot)$ indicates momentum contractions and the numerical coefficients denote summation over different channels. Given the GR prediction for the parity-insensitivity of $A,B$, due to the absence of parity-violating scalar quadratic/cubic functions, $\psi_{\varphi \varphi}$ and $\psi_{\varphi \varphi \varphi}$ are non-perturbatively parity-even and therefore drop out of the parity-odd sector. Thus we are left with a trivial rescaling,
\begin{align}
	\delta\psi_{\zeta \zeta \zeta \zeta}=\left(\frac{\dot{\phi}_0}{H}\right)^4 \delta\psi_{\varphi \varphi \varphi \varphi}~.
\end{align}
Combining everything, we finally arrive at the leading-order prediction for the curvature trispectrum,
\begin{keyeqn}
	\begin{align} 
		B_{\zeta \zeta \zeta \zeta}^{\text{PO}}  
		=&-i\left(\frac{H}{\dot{\phi}_0}\right)^4\,\frac{\pi\kappa H^7}{2M^2_{\rm{pl}}}\frac{[\hat{\bfk}_1\cdot\hat{\bfk}_3-(\hat{\bfk}_{1}\cdot\hat{\bf{s}})(\hat{\bfk}_{3}\cdot\hat{\bf{s}})]}{(k_1 k_2 k_3 k_4)^2}\frac{\hat{\bf{s}}\cdot(\hat{\bfk}_2\times\hat{\bfk}_4)}{s^3 E^2_L E^2_R} \left(E_L^3-E_L\,e_2-e_3\right) \left(E_R^3-E_R\,\tilde{e}_2-\tilde{e}_3\right)\nonumber\\
		&+\text{2 perms.}\label{curvatureTrispectrum}
	\end{align}
\end{keyeqn}
The GR graviton power spectrum and its first-order correction are
\begin{keyeqn}
	\begin{align}
            B^{(h)}_{\gamma \gamma}(s) =\frac{H^2}{M^2_{\rm{pl}}s^3}~,\quad \delta B^{(h)}_{\gamma \gamma}(s) =\frac{h\pi\kappa H^3}{M^2_{\rm{pl}}s^3}~,\label{gravitonPowerspectrum}
	\end{align}
\end{keyeqn}
along with the well-celebrated curvature-graviton mixed bispectrum à la Maldacena \cite{Maldacena:2002vr},
\begin{keyeqn}
	\begin{align}
		B^{(h)}_{\zeta\zeta \gamma}(\bfk_1,\bfk_2,-{\bf{s}}) =\frac{H^6}{2M_{\rm{pl}}^2\,\dot{\phi}_0^2}\frac{k_{1i}\,k_{2j}\,\mathrm{e}^{(h)}_{ij}(\mathbf{s})}{k_1^3k_2^3s^3}\times \left[\frac{1}{E_L^2}\left(-E_L^3+E_L\,e_2+e_3\right)\right]\,.\label{gravitonBispectrum}
	\end{align}
\end{keyeqn}
Using the polarisation identity (\ref{PolarSum}), it is almost a trivial task to verify that \eqref{curvatureTrispectrum}, \eqref{gravitonPowerspectrum} and \eqref{gravitonBispectrum} do indeed satisfy our CCF relation \eqref{CSCCF}. 

\begin{figure}[h!]
    \centering
    \hspace{-1.5cm}
    \includegraphics[width=0.55\linewidth]{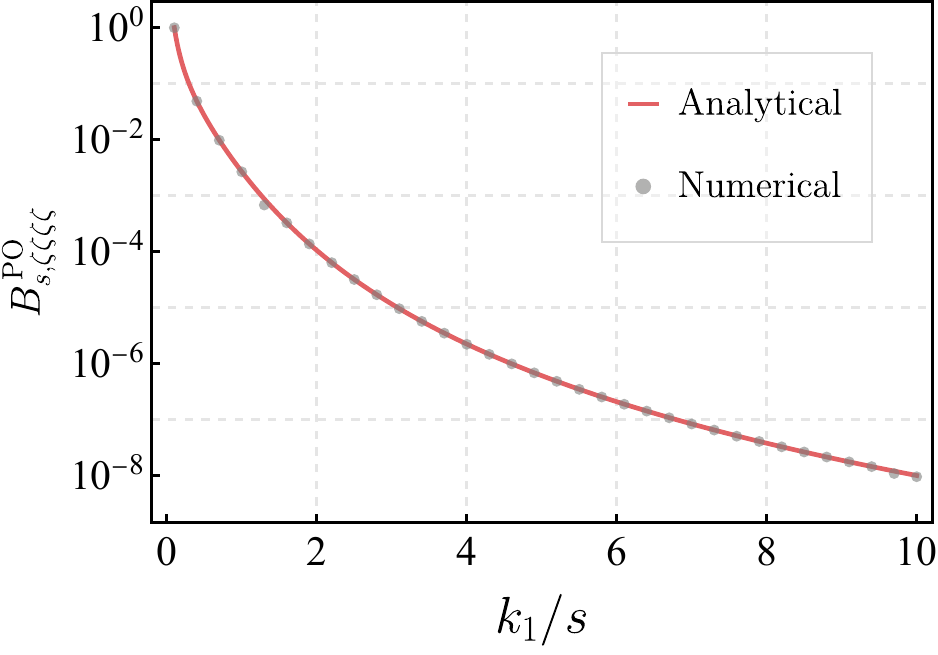}
    \caption{The $s$-channel parity-odd trispectrum $B_{s,\zeta \zeta \zeta \zeta}^{\text{PO}}$ as function of the momentum ratio $k_1/s$. The \textcolor[RGB]{226,97,100}{red solid line} shows the analytical result obtained from our scalar-tensor CCF formula \eqref{curvatureTrispectrum},
    and \textcolor[RGB]{178,179,178}{gray dots} represent the  numerical results from a direct, brute-force calculation of the complete time integrals of the full trispectrum. We see that the two methods match perfectly. For better visualisation, the overall size is normalised to its maximum value. We set the tetrahedron kinematics configuration as $k_1=k_3,\,k_2=k_4=\sqrt{s^2+k_1^2}$, and $\psi=\pi/3$ being the dihedral angle from $(\bfk_1,\bfk_2)$-plane to $(\bfk_3,\bfk_4)$-plane.}\label{Fig:Num}
\end{figure}

In Figure \ref{Fig:Num}, we plot the parity-odd trispectrum $B^{\rm{PO}}_{\zeta\zeta\zeta\zeta}$ from Chern-Simons gravity, for both the simple analytical expression we have found in this work given in \eqref{curvatureTrispectrum} and a brute-force numerical calculation where we compute the necessary time integrals of the in-in/Schwinger-Keldysh formalisms. The close agreement between the two clearly verifies the correctness of our scalar-tensor CCF formula. In Appendix \ref{Sec:appA}, we provide additional details on how we performed the numerical integration, and we present an alternative derivation of the CCF expression \eqref{curvatureTrispectrum} using the traditional Schwinger-Keldysh formalism. We also present a comparison between our result and that of \cite{Creque-Sarbinowski:2023wmb}.


\section{A double-copy of graviton bispectra} \label{sec5}

We have so far established a concrete consistency relation among inflationary observables. Interestingly, to linear order in $f^{-1}$, the double-cut CCF relation \eqref{CSCCF} can be packed into a neat single-cut form.\footnote{Here the number of ``cuts" is counting the number of power spectra in the CCF denominator.} To achieve this, we combine the GR graviton power spectrum and its correction into the total power spectrum,
\begin{align}
	\widehat B^{(h)}_{\gamma \gamma}\equiv B^{(h)}_{\gamma \gamma}+\delta B^{(h)}_{\gamma \gamma}~,
\end{align}
and introduce the total curvature-graviton mixed bispectrum by
\begin{align}
    \widehat B^{(h)}_{\zeta\zeta \gamma}\equiv B^{(h)}_{\zeta\zeta \gamma}\left(1+\frac{\delta B^{(h)}_{\gamma \gamma}}{B^{(h)}_{\gamma \gamma}}\right)~.
\end{align}
This leads to
\begin{keyeqn}
\begin{align} \label{CSCCFFull}
	B_{\zeta\zeta\zeta\zeta}^{\text{PO}} =  \sum_{h = \pm} \left[\frac{\widehat B^{(h)}_{\zeta \zeta \gamma}(\bfk_1, \bfk_2, -\mathbf{s}) \, \widehat B^{(h)}_{\zeta \zeta \gamma}(\bfk_3, \bfk_4, \mathbf{s})}{\widehat B^{(h)}_{\gamma \gamma}(s)}\right]^{\rm PO} + 2 ~ \text{perms} \,.
\end{align}
\end{keyeqn}
Here the projection onto the parity-odd sector $[\cdots]^{\rm PO}$ singles out the term linear in $\delta B^{(h)}_{\gamma \gamma}$ in the expansion of the reciprocal power spectrum and \eqref{CSCCFFull} is therefore equivalent to \eqref{CSCCF} at leading order in $f^{-1}$. This form of the CCF relation states that the parity-odd curvature trispectrum is given by a \textit{double-copy} of the graviton-curvature mixed bispectrum, as long as the Chern-Simons term in \eqref{definingAction} is the sole source of parity violation. A few comments follow from this scalar-tensor CCF relation:
\begin{itemize}
	\item We stress that \eqref{CSCCFFull} is an immediate \textit{prediction} of the CCF formula in \cite{Stefanyszyn:2024msm} when its assumptions of validity has been checked. The essential ingredient is the IR convergence of the connected wavefunctions as shown in Section \ref{sec4}. Once this is done, we can directly deduce the scalar-tensor CCF relation \eqref{CSCCFFull} without explicitly computing each of its ingredients (which we nevertheless did in Section \ref{sec4} only to stay explicit).
	\item The scalar-tensor CCF relation technically follows from the reality theorem of \cite{Stefanyszyn:2023qov}, and thereby inherits its premises. Most importantly, we require exact scale invariance, Bunch-Davies vacuum state and the tree approximation. It also assumes general principles like unitarity and locality. Breaking any of these premises can (but not always) lead to deviation from the scalar-tensor CCF relation \eqref{CSCCFFull}.
	\item Gravity is notoriously weak and therefore the reader will not be surprised that the signal strength of the parity-odd trispectrum generated due to the mixing with gravity is feebly weak at linear order in $\kappa$. The size estimator of the parity-odd trispectrum is roughly given by
    \begin{align}
        g_{\rm NL}^{\rm PO}\sim \frac{B_{\zeta\zeta\zeta\zeta}^{\text{PO}}}{B_{\zeta\zeta}B_{\zeta\zeta}B_{\zeta\zeta}}\sim \frac{\dot{\phi}_0^2}{H^4}\times\frac{\kappa H^3}{M_{\rm pl}^2}\sim \kappa H \varepsilon\lesssim \varepsilon \,,
    \end{align}
    which is beyond the scope of current observations. This size is comparable to that of the parity-even trispectrum \cite{Seery:2006vu,Seery:2008ax}. However, we note that there exist scenarios in which the graviton non-Gaussianities are greatly enhanced \cite{Dimastrogiovanni:2014ina,Dimastrogiovanni:2015pla}. It is interesting to study whether similar CCF relations hold there as well. We leave such investigations to future work.
\end{itemize} 

\subsubsection*{Beyond Chern-Simons}

Following the common wisdom of EFTs, the Chern-Simons operator in \eqref{definingAction} must appear as one of the higher-dimensional operators ($O_i$'s) in the infinite tower of effective interactions,
\begin{align}
    S_{\rm EFT}= \int \d^4 x \sqrt{-g}\left[\frac{M^2_{\rm{pl}}}{2}R -\frac{1}{2}g^{\mu\nu}\partial_\mu \phi \partial_\nu \phi - V(\phi)+\sum_{i=1}^\infty c_i O_i\right]~,\label{generalEFTs}
\end{align}
where $c_i$'s are the Wilson coefficients whose values define a specific EFT. In fact, the Chern-Simons term is the leading operator that violates parity in the gravity sector with a metric formulation. There is no intrinsic reason why it should be absent. What matters is merely the size of its Wilson coefficient ($c_{\rm CS}\propto f^{-1}$). Yet in more general settings, the gravitational Chern-Simons term might not be the dominant parity-violating operator in the infinite EFT tower. For instance, one can introduce other spinning fields such as vector bosons ($A$), fermions ($\psi$) or even higher-spin fields ($\Sigma$) whose interactions in $\{O\}$ violate parity. Based upon the high mass-dimensionality of these parity-violating interactions (as discussed in e.g. \cite{Stefanyszyn:2023qov}), it is not unnatural to assume IR convergence of the associated time integrals in which case our CCF predicts:
\begin{align}
	B_{\zeta\zeta\zeta\zeta}^{\text{PO}} =  3\left[B_{\zeta \zeta \gamma}\cdot \frac{1}{\widehat B_{\gamma \gamma}}\cdot B_{\zeta \zeta \gamma}\right]^{\rm PO} +\sum_{X=A,\Sigma,\cdots} 3\left[B_{\zeta \zeta X}\cdot \frac{1}{\widehat B_{XX}}\cdot B_{\zeta \zeta X}\right]^{\rm PO}+\text{fermions}+\cdots~,\label{CCFgenerallyCorrected}
\end{align}
where $3$ stands for summing over channels and $X$ traverses the whole bosonic spectrum. The first term here is from the visible sector i.e. the graviton, whereas all other terms are in the invisible/dark sector and cannot be probed by current measurements (see, however, \cite{Li:2020xwr,Lu:2021gso}). The ellipsis here denotes all other sources of parity violation unaccounted for by the CCF formula. In general, each term in \eqref{CCFgenerallyCorrected} originates from different sources and they are therefore not correlated with each other. Therefore, despite their signs being indefinite, we expect them to add up on average, meaning that within an ensemble of EFTs, we have
\begin{align}
    \overline{~\left\|B_{\zeta\zeta\zeta\zeta}^{\text{PO}}\right \|^2~}=3\overline{~\left\|\left[B_{\zeta \zeta \gamma}\cdot \frac{1}{\widehat B_{\gamma \gamma}}\cdot B_{\zeta \zeta \gamma}\right]^{\rm PO}\right\|^2~} +\sum_{X=A,\Sigma,\cdots} 3\overline{~\left\|\left[B_{\zeta \zeta X}\cdot \frac{1}{\widehat B_{XX}}\cdot B_{\zeta \zeta X}\right]^{\rm PO}\right\|^2~}+\cdots~,
\end{align}
where $\|\cdot\|$ denotes a norm with respect to some positive-definite inner product defined in kinematic space,\footnote{For bispectra, this is often known by the cosine \cite{Babich:2004gb} and its generalisations \cite{Fergusson:2008ra}.} and
\begin{align}
    \overline{B}\equiv  \frac{\int\prod_i \d c_i \,\varrho(\{c\}) B(\{c\})}{\int \prod_i \d c_i \,\varrho(\{c\})} \,,
\end{align}
denotes an average over the ensemble of EFTs with Wilson coefficients $\{c\}$ that is weighed by some probability density function $\varrho(\{c\})$. Note that the cross terms average to zero since they are uncorrelated in the general ensemble of EFTs. Positivity of the contribution from the invisible/dark sector thus yields an inequality among visible observables:
\begin{keyeqn}
    \begin{align}
    \overline{~\left\|B_{\zeta\zeta\zeta\zeta}^{\text{PO}}\right \|^2~}\geq 3\overline{~\left\|\left[B_{\zeta \zeta \gamma}\cdot \frac{1}{\widehat B_{\gamma \gamma}}\cdot B_{\zeta \zeta \gamma}\right]^{\rm PO}\right\|^2~} ~. \label{CCFEFTEnsembleInequality}
\end{align}
\end{keyeqn}
This inequality saturates whenever the assumptions of the CCF are satisfied and the Chern-Simons operator is the sole source of parity violation.

Of course, the EFT ensemble average in \eqref{CCFEFTEnsembleInequality} cannot be performed in reality. Therefore, the practical message conveyed here is that we expect the size of the observed parity-odd trispectrum to naturally be greater than or equal to the double-copied graviton bispectra. If observations indicated that
\begin{align}
    \left\|B_{\zeta\zeta\zeta\zeta}^{\text{PO}}\right \|^2\neq  3\left\|\left[B_{\zeta \zeta \gamma}\cdot \frac{1}{\widehat B_{\gamma \gamma}}\cdot B_{\zeta \zeta \gamma}\right]^{\rm PO}\right\|^2 ~,
\end{align}
the interpretation is that there could exist extra sources of parity violation in the invisible/dark sector, whereas a clear violation of \eqref{CCFEFTEnsembleInequality},
\begin{align}
    \left\|B_{\zeta\zeta\zeta\zeta}^{\text{PO}}\right \|^2\ll 3\left\|\left[B_{\zeta \zeta \gamma}\cdot \frac{1}{\widehat B_{\gamma \gamma}}\cdot B_{\zeta \zeta \gamma}\right]^{\rm PO}\right\|^2 ~,\label{CCFEFTEnsembleInequalityViolation}
\end{align}
would indicate that the CCF is falsified by observation and at least one of its assumptions must be broken. More likely, either scale invariance or the Bunch-Davies initial condition must be badly broken, or that there exist large quantum corrections from loop effects. Less likely causes are the loss of unitarity and locality in the quantum field theory of inflation. Many of these possibilities can be verified independently in other observables. To name a few examples, the breaking of scale invariance is often manifest in all correlators, while the non-Bunch-Davies initial conditions can be read off of the non-Gaussianities in the colinear/folded kinematic limits. In this sense, \eqref{CCFEFTEnsembleInequality} (or more precisely, the absence of its violation, \eqref{CCFEFTEnsembleInequalityViolation}) serves as a \textit{null test} that verifies the fundamental principles of physics during inflation.\footnote{We thank Gilles Esposito-Farèse for suggesting this perspective.}


\section{Conclusions} \label{sec6}

Observables play a central role in modern physics. Exploring their internal structure and mutual relationships often bring some of the deepest insights on the fundamental principles of Nature. In this work, we have indeed found such a profound relationship in the context of primordial cosmology. We have shown that in the minimal model of inflation that violates parity, where a gravitational Chern-Simons term is included, the resulting parity-odd curvature trispectrum is a simple ``double-copy'' of curvature-graviton bispectra for all kinematics. Such a scalar-tensor CCF relation derives directly from the principles of unitarity, locality, scale invariance, Bunch-Davies vacuum and classicality. It serves as a first explicit example of a cosmological cutting rule that can be directly tested by future observations. Observing any significant deviation would point straight to the failure of at least one of the underlying assumptions and therefore constitutes as a null test of fundamental principles involved during inflation.

\vskip 5pt
There are a few interesting directions for future research:
\begin{itemize}
    \item Due to the Planckian suppression of the gravitational coupling, the signal strength in our minimal model is too weak to be probed by current observations. However, our minimal model serves merely as a first step towards testing relations in observably large correlators. Going beyond the minimal setup, there are plenty of models producing large graviton non-Gaussianities \cite{Dimastrogiovanni:2014ina,Dimastrogiovanni:2015pla,Bartolo:2018elp,Orlando:2025pkb,Tong:2022cdz}. One can then try to examine whether a similar scalar-tensor CCF exists in these models.
    \item In this work, we have only worked up to linear order in the Chern-Simons coupling $f^{-1}$, which is valid as long as the chemical potential is small, $\kappa H\lesssim 1$. In the case where $\kappa H>1$, one needs to resum the infinite chain of linear mixings in the graviton linear theory. It is conceivable that this can bring an exponential enhancement of the graviton. However, the ghost instability problem reappears after resummation. To fully investigate the $\kappa H>1$ case, we need to find a partial UV completion that solves the ghost instability. 
    \item Moving beyond the metric formulation of gravity, one can introduce a Nieh-Yan term in teleparallel gravity to source parity violation \cite{Nieh:1981ww,Li:2020xjt}. It has been shown that the Nieh-Yan term can lead to large chiral primordial gravitational waves, and parity-violating non-Gaussianities \cite{Cai:2021uup,Rao:2023doc,Xu:2024kwy,Wu:2021ndf,Zhang:2024vfw}. Other scenarios that generate larger graviton parity violation signals include chiral scalar-tensor theories \cite{Moretti:2024fzb}, alternative to inflationary scenarios \cite{Cai:2022lec,Akama:2024bav,Zhu:2023lhv} and general initial conditions\cite{Gong:2023kpe,Christodoulidis:2024ric}. It is therefore interesting to examine whether a similar scalar-tensor CCF applies in those cases as well.
    \item Despite the common belief that curvature and graviton modes are the only measurable fields at the end of inflation, other light degrees of freedom (such as an axion) may also be measured as isocurvature perturbations, at least in principle. While no significant evidence has been found so far for isocurvature modes, their correlators are valid observables after all. It is thus interesting to inspect the CCF relations involving isocurvature correlators and construct models realising such ``isocurvature colliders'' \cite{Li:2020xwr,Lu:2021gso}.
\end{itemize}


\paragraph*{Acknowledgements} We thank Shingo Akama, Trevor Cheung, Cyril Creque-Sarbinowski, Zongzhe Du, Gilles Esposito-Farèse,  Chunshan Lin, Giorgio Orlando, Enrico Pajer and Masahide Yamaguchi for helpful discussions. D.S. is supported by a UKRI Stephen Hawking Fellowship [grant number EP/W005441/1] and a Nottingham Research Fellowship from the University of Nottingham. X.T. is supported by STFC consolidated grants ST/T000694/1 and ST/X000664/1. Y.Z. is supported by the IBS under the project code, IBS-R018-D3. X.T. and Y.Z. thank the University of Nottingham for kind hospitality. For the purpose of open access, the authors have applied a CC BY public copyright licence to any Author Accepted Manuscript version arising.

\appendix

\section{Correlator-to-correlator factorisation from the Schwinger-Keldysh formalism }\label{Sec:appA}

In this appendix, for readers unfamiliar with the wavefunction formalism in the main text, we provide an alternative derivation of our scalar-tensor CCF result using the more conventional Schwinger-Keldysh (SK) formalism. As stated in Section \ref{sec4}, Contributions II and III do not yield any parity-odd contribution, hence we will only focus on calculating Contribution I here. Using the cubic interaction (\ref{3ptint}) and the parity-odd vertex from the Chern-Simons correction (\ref{2ptint}) and following the diagrammatic rules in \cite{Chen:2017ryl}, we have 
\begin{align}
    B^{\rm{PO}}_{\varphi\varphi\varphi\varphi} &= \raisebox{-13pt}{\includegraphics[scale=0.6]{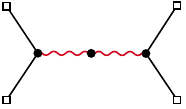}}~+~\raisebox{-13pt}{\includegraphics[scale=0.6]{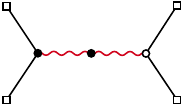}}~+~\raisebox{-13pt}{\includegraphics[scale=0.6]{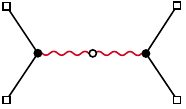}}~+~\raisebox{-13pt}{\includegraphics[scale=0.6]{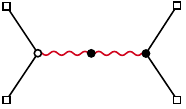}}\nonumber\\[0.5em]
    &+\raisebox{-13pt}{\includegraphics[scale=0.6]{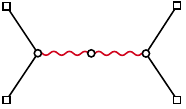}}~+~\raisebox{-13pt}{\includegraphics[scale=0.6]{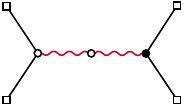}}~+~\raisebox{-13pt}{\includegraphics[scale=0.6]{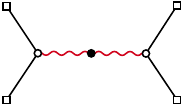}}~+~\raisebox{-13pt}{\includegraphics[scale=0.6]{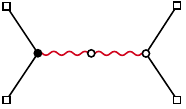}}\,. \label{B4_SKdiagramsAll}
\end{align}
The diagrams in the SK formalism differ from those used in the main text for wavefunction coefficients. Here, the white square $\Box$ denotes the time boundary $\eta_0$, at which we evaluate the correlators. Each vertex has two possible colour assignments to indicate whether they are from the time-ordered path (labelled by black $\bullet$ or ``$+$'') or the anti-time-ordered (labelled by white $\circ$ or ``$-$'') one. So for the three-vertex topology, we have in total $2^3=8$ coloured diagrams. The black and white vertices are related by complex conjugation and the reversal of spatial momenta at that vertex. We caution the readers that the flip of momenta may have often been overlooked in the literature. While this momentum reversal matters not in parity-conserving theories, it becomes crucial when the theory is parity-violating. For example,
\begin{align}
    \Bigg(~\raisebox{-14pt}{\includegraphics[scale=0.6]{Fig/SK_bbb.pdf}}~\Bigg)_{\bfk}\quad=\quad \Bigg(~\raisebox{-14pt}{\includegraphics[scale=0.6]{Fig/SK_www.pdf}}~\Bigg)^*_{\bfk\rightarrow -\bfk}\overset{\text{PO}}{=}\quad-\Bigg(~\raisebox{-14pt}{\includegraphics[scale=0.6]{Fig/SK_www.pdf}}~\Bigg)^*_{\bfk}\,.
\end{align}
Thus, the second line in (\ref{B4_SKdiagramsAll}) is the complex conjugate of the first line with an additional sign flip, resulting in $B^{\rm{PO}}_{\varphi\varphi\varphi\varphi}$ being purely imaginary (as it should be):
\begin{align}
    B^{\rm{PO}}_{\varphi\varphi\varphi\varphi} &= \raisebox{-13pt}{\includegraphics[scale=0.6]{Fig/SK_bbb.pdf}}~+~\raisebox{-13pt}{\includegraphics[scale=0.6]{Fig/SK_bbw.pdf}}~+~\raisebox{-13pt}{\includegraphics[scale=0.6]{Fig/SK_bwb.pdf}}~+~\raisebox{-13pt}{\includegraphics[scale=0.6]{Fig/SK_wbb.pdf}}-\text{c.c.}\,,\label{B4_SKdiagrams}
\end{align}
One can spell \eqref{B4_SKdiagramsAll} out more explicitly using the SK Feynman rules,
\begin{align}
    B^{\rm{PO}}_{\varphi\varphi\varphi\varphi}(\bfk_1,\bfk_2,\bfk_3,\bfk_4)&= i\,\frac{2 M^2_{\text{pl}}\kappa s}{H^5}\sum_{\sf{a,b,c}=\pm}\sum_{h = \pm} h\,\Big(\bfk_{1}\cdot\hat{\mathbf{e}}^{(h)}(\mathbf{s})\Big)^*\Big(\bfk_{2}\cdot\hat{\mathbf{e}}^{(h)}(\mathbf{s})\Big)^*\Big(\bfk_{3}\cdot\hat{\mathbf{e}}^{(h)}(\mathbf{s})\Big)\Big(\bfk_{4}\cdot\hat{\mathbf{e}}^{(h)}(\mathbf{s})\Big)\nonumber\\[0.3em]
	&~~~\times \,{\sf{abc\,}}\int\frac{d\eta_1}{\eta_1^2}\frac{d\eta_2}{\eta_2}\frac{d\eta_3}{\eta_3^2}\mathcal{K}^{\varphi}_{\sf{a}}(\eta_1,k_1)\mathcal{K}^{\varphi}_{\sf{a}}(\eta_1,k_2)\mathcal{K}^{\varphi}_{\sf{c}}(\eta_3,k_3)\mathcal{K}^{\varphi}_{\sf{c}}(\eta_3,k_4)\nonumber
	\\[0.3em]
    &~~~\times\left[\partial_{\eta_2}\mathcal{G}^\gamma_{\sf{ab}}(\eta_1,\eta_2,s)\partial_{\eta_2}\mathcal{G}^\gamma_{\sf{bc}}(\eta_2,\eta_3,s)-s^2 \mathcal{G}^\gamma_{\sf{ab}}(\eta_1,\eta_2,s)\mathcal{G}^\gamma_{\sf{bc}}(\eta_2,\eta_3,s)\right] \nonumber \\[0.3em] &~~~+\text{2 perms}~.\,\label{B4_SKexpression}
\end{align}
Letters in \textsf{sans serif} font $(\sf{a, b, \ldots})$ here denote SK indices. To avoid confusion with the wavefunction notation, SK propagators are written in the calligraphic font. The external propagators are given by,
\begin{subequations}
\begin{align}
    &\mathcal{K}^{\varphi}_+(k,\eta)=\varphi(\eta_0,k)\varphi^*(\eta,k)\,,\qquad \mathcal{K}^{\gamma}_+(k,\eta)=\gamma_h(\eta_0,k)\gamma_h^*(\eta,k)\,,\\
    &\mathcal{K}^{\varphi}_-(k,\eta)=\varphi^*(\eta_0,k)\varphi(\eta,k)\,,\qquad \mathcal{K}^{\gamma}_+(k,\eta)=\gamma_h^*(\eta_0,k)\gamma_h(\eta,k)\,,
\end{align}
\end{subequations}
where the mode functions of the massless inflaton and graviton fields are given by
\begin{align}
    \varphi(\eta,k)=\frac{M_{\rm{pl}}}{2}\gamma_h(\eta,k)=\frac{H}{\sqrt{2k^3}}(1+ik\eta) e^{-i k\eta}\,.
\end{align}
The internal propagators connecting two different vertices read
\begin{subequations}
	\begin{align}
		\mathcal{G}^\gamma_{++}(\eta_1,\eta_2,k)&=\theta(\eta_1-\eta_2)\gamma(\eta_1,k)\gamma^*(\eta_2,k)+\theta(\eta_2-\eta_1)\gamma^*(\eta_1,k)\gamma(\eta_2,k)~,\\
		\mathcal{G}^\gamma_{+-}(\eta_1,\eta_2,k)&=\gamma^*(\eta_1,k)\gamma(\eta_2,k)~,\\
		\mathcal{G}^\gamma_{-+}(\eta_1,\eta_2,k)&=\gamma(\eta_1,k)\gamma^*(\eta_2,k)~,\\ 
            \mathcal{G}^\gamma_{--}(\eta_1,\eta_2,k)&=\theta(\eta_1-\eta_2)\gamma^*(\eta_1,k)\gamma(\eta_2,k)+\theta(\eta_2-\eta_1)\gamma(\eta_1,k)\gamma^*(\eta_2,k)~.
	\end{align}
\end{subequations}
Unlike the wavefunction coefficients $\delta\psi_{\varphi\varphi\varphi\varphi}, \delta \psi_{\varphi\varphi\gamma}$ discussed in Section \ref{sec4} that converge at late times, each individual SK diagrams in (\ref{B4_SKdiagrams}) posses \textit{intrinsic} IR divergence. As an example, consider the IR scaling behaviour of the fully factorised diagram,
\begin{align}
    \raisebox{-13pt}{\includegraphics[scale=0.6]{Fig/SK_bwb.pdf}}&\sim \int^{\eta_0}\frac{d\eta_1}{\eta_1^2}\frac{d\eta_2}{\eta_2}\frac{d\eta_3}{\eta_3^2}\mathcal{K}^{\varphi}_{+}(\eta_1,k_1)\mathcal{K}^{\varphi}_{+}(\eta_1,k_2)\mathcal{K}^{\varphi}_{+}(\eta_3,k_3)\mathcal{K}^{\varphi}_{+}(\eta_3,k_4)\nonumber
	\\[0.3em]
    &\times\left[\partial_{\eta_2}\mathcal{G}^\gamma_{+-}(\eta_1,\eta_2,s)\partial_{\eta_2}\mathcal{G}^\gamma_{-+}(\eta_2,\eta_3,s)-s^2 \mathcal{G}^\gamma_{+-}(\eta_1,\eta_2,s)\mathcal{G}^\gamma_{-+}(\eta_2,\eta_3,s)\right]\nonumber\\
    &\sim\int^{\eta_0} \frac{d\eta_1}{\eta_1^2}\mathcal{O}(\eta_1^0)\int^{\eta_0}\frac{d\eta_2}
    {\eta_2}\mathcal{O}(\eta_2^0)\int^{\eta_0}\frac{d\eta_3}{\eta_3^2}\mathcal{O}(\eta_3^0)\nonumber\\
    &\sim{O}\left(\eta_0^{-2}\log \eta_0\right)\,,
\end{align}
which has severe IR divergence, as both the external and internal propagators scale as $\mathcal{O}(\eta^0)$ in the IR limit. It is only after summing all SK diagrams that divergences such as this can eventually cancel out. Due to this complication, the SK computation is extremely subtle. Any improper approximation or neglecting any tiny pieces of contribution can lead to the failure of this fine cancellation and thus a wrong result. Inspired by the bulk-bulk wavefunction propagator, and to apply the reality theorem in hope of solving the IR complication, we decompose the Feynman propagator into two parts i.e. $\mathcal{G}_{++}=C_{++} + F_{++}$, following the method proposed in \cite{Stefanyszyn:2023qov}:
\begin{align}
    &C_{++}=\mathcal{G}(\eta_1,\eta_2,k)-\gamma^*_{h}(\eta_1,k)\gamma^*_h(\eta_2,k)\,,\\
    &F_{++}=\gamma^*_{h}(\eta_1,k)\gamma^*_h(\eta_2,k)\,,
\end{align}
or diagrammatically,
\begin{align}
    \raisebox{-3pt}{\includegraphics[scale=1]{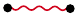}}\quad=\quad\raisebox{-3pt}{\includegraphics[scale=1]{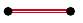}}\quad+\quad\raisebox{-3pt}{\includegraphics[scale=1]{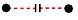}}\,,
\end{align}
where the double line represents the \textit{connected} propagator $C_{++}$ which includes all the time-orderings, while the dashed line denotes the \textit{factorised} propagator $F_{++}$. Comparing with (\ref{bulktobulk}), one observes that $C_{++}$ is nothing else but the bulk-bulk propagator, and thus has the softened IR scaling (\ref{bulkIR}) i.e. $C_{++}=G\sim\eta^3$. Now we can perform the decomposition for all Feynman propagators in (\ref{B4_SKdiagrams}). Some contributions, for example,
\begin{align}
    \raisebox{-13pt}{\includegraphics[scale=0.6]{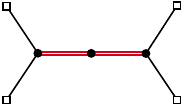}}~~\quad\quad\raisebox{-13pt}{\includegraphics[scale=0.6]{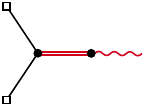}}\quad\quad\raisebox{-13pt}{\includegraphics[scale=0.6]{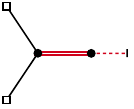}}~~,
\end{align}
become IR finite, following the same reasoning for $\delta\psi_{\varphi\varphi\varphi\varphi}$ and $\delta\psi_{\varphi\varphi\gamma}$. As a result, the reality theorem can also be applied, from which it follows immediately that these parts drop out of the parity-odd correlator. Then we have
\begin{align}
    B^{\rm{PO}}_{\varphi\varphi\varphi\varphi} &= \raisebox{-13pt}{\includegraphics[scale=0.6]{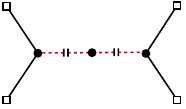}}~+~\raisebox{-13pt}{\includegraphics[scale=0.6]{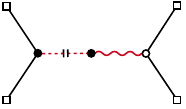}}~+~\raisebox{-13pt}{\includegraphics[scale=0.6]{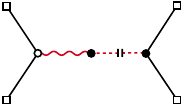}}~+~\raisebox{-13pt}{\includegraphics[scale=0.6]{Fig/SK_bwb.pdf}}-\text{c.c.}\,.\label{B4_SKCFdecomp}
\end{align}
Notice that $\mathcal{G}_{\pm\mp}$ and $F_{+-}$ are related to the external propagator through 
\begin{align}
    &\mathcal{G}^\gamma_{\pm\mp}(k,\eta_1,\eta_2)=\frac{1}{B_{\gamma\gamma}} \mathcal{K}_{\pm}(k,\eta_1) \mathcal{K}_{\mp}(k,\eta_2)\,,\\
    &F_{++}(k,\eta_1,\eta_2)=\frac{1}{B_{\gamma\gamma}} \mathcal{K}_{+}(k,\eta_1) \mathcal{K}_{+}(k,\eta_2)\,,
\end{align}
then we can write (\ref{B4_SKCFdecomp}) in a different way by replacing internal propagators to the external one with the power spectrum,
\begin{align}
     B^{\rm{PO}}_{\varphi\varphi\varphi\varphi}~=&\frac{1}{\left(\raisebox{0pt}{\,\includegraphics[scale=0.5]{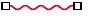}}\right)^2}\Bigg[~\raisebox{-11pt}{\includegraphics[scale=0.5]{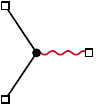}}~\raisebox{0pt}{\includegraphics[scale=0.5]{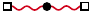}}~\raisebox{-11pt}{\includegraphics[scale=0.5]{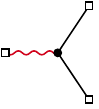}}+\raisebox{-11pt}{\includegraphics[scale=0.5]{Fig/SK_B3Lb.pdf}}~\raisebox{0pt}{\includegraphics[scale=0.5]{Fig/SK_hB2b.pdf}}~\raisebox{-11pt}{\includegraphics[scale=0.5]{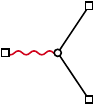}}+\raisebox{-11pt}{\includegraphics[scale=0.5]{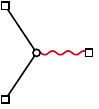}}~\raisebox{0pt}{\includegraphics[scale=0.5]{Fig/SK_hB2b.pdf}}~\raisebox{-11pt}{\includegraphics[scale=0.5]{Fig/SK_B3Rb.pdf}}+\raisebox{-11pt}{\includegraphics[scale=0.5]{Fig/SK_B3Lb.pdf}}~\raisebox{0pt}{\includegraphics[scale=0.5]{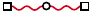}}~\raisebox{-11pt}{\includegraphics[scale=0.5]{Fig/SK_B3Rb.pdf}}\nonumber\\
     &~~~~~~~~~~+\raisebox{-11pt}{\includegraphics[scale=0.5]{Fig/SK_B3Lw.pdf}}~\raisebox{0pt}{\includegraphics[scale=0.5]{Fig/SK_hB2w.pdf}}~\raisebox{-11pt}{\includegraphics[scale=0.5]{Fig/SK_B3Rw.pdf}}+\raisebox{-11pt}{\includegraphics[scale=0.5]{Fig/SK_B3Lw.pdf}}~\raisebox{0pt}{\includegraphics[scale=0.5]{Fig/SK_hB2w.pdf}}~\raisebox{-11pt}{\includegraphics[scale=0.5]{Fig/SK_B3Rb.pdf}}+\raisebox{-11pt}{\includegraphics[scale=0.5]{Fig/SK_B3Lb.pdf}}~\raisebox{0pt}{\includegraphics[scale=0.5]{Fig/SK_hB2w.pdf}}~\raisebox{-11pt}{\includegraphics[scale=0.5]{Fig/SK_B3Rw.pdf}}+\raisebox{-11pt}{\includegraphics[scale=0.5]{Fig/SK_B3Lw.pdf}}~\raisebox{0pt}{\includegraphics[scale=0.5]{Fig/SK_hB2b.pdf}}~\raisebox{-11pt}{\includegraphics[scale=0.5]{Fig/SK_B3Rw.pdf}}~\Bigg]\,.
\end{align}
After a simple recombination of terms, we arrive at
\begin{align}
    B^{\rm{PO}}_{\varphi\varphi\varphi\varphi}~&=\left(~\raisebox{-11pt}{\includegraphics[scale=0.5]{Fig/SK_B3Lb.pdf}}\,+\raisebox{-11pt}{\includegraphics[scale=0.5]{Fig/SK_B3Lw.pdf}}\right)~\frac{\big(\raisebox{0pt}{\includegraphics[scale=0.5]{Fig/SK_hB2b.pdf}}+\raisebox{0pt}{\includegraphics[scale=0.5]{Fig/SK_hB2w.pdf}}\big)}{~\left(\raisebox{0pt}{\,\includegraphics[scale=0.5]{Fig/SK_B2gamma.pdf}}\right)^2}~\left(~\raisebox{-11pt}{\includegraphics[scale=0.5]{Fig/SK_B3Rb.pdf}}+\,\raisebox{-11pt}{\includegraphics[scale=0.5]{Fig/SK_B3Rw.pdf}}\right)\nonumber\\
    &=\sum_{h = \pm} B^{(h)}_{\varphi \varphi \gamma}\,\frac{ \delta B^{(h)}_{\gamma \gamma}(s) }{B^{(h)}_{\gamma \gamma}(s) \, B^{(h)}_{\gamma \gamma}(s)}\, B^{(h)}_{\varphi \varphi \gamma}\,,
\end{align}
which is exactly the scalar-tensor CCF formula \eqref{CSCCF} we found in our main text.

\subsubsection*{Numerical calculation}
\begin{figure}[h!]
	\centering
	\begin{tikzpicture}
		[scale = 1.5]
		\draw[->,line width=0.8pt] (0, -1.0) -- (0, 1.5) coordinate[label=below:$ $] (k1);
		\draw[->,line width=0.8pt] (-1.5, 0) -- (1.5, 0) coordinate[label=below:$ $] (k1);
		\node at (0, 1.7) {$\text{Im}(\eta)$};
		\node at (1.9, 0) {$\text{Re}(\eta)$};
            \node at (-1.0, -0.2) {$\eta_*$};
		
		\path[blue2, draw, line width = 1.2pt, postaction = decorate, decoration={markings, mark=at position 0.5 with {\arrow[line width=1pt]{>}}}] (-1.0, 1.6) -- (-1.0, 0);
		\node at (-0.75, 0.8) {\textcolor{blue2}{$\mathcal{C}_1$}};
		
		\path[green2, draw, line width = 1.2pt, postaction = decorate, decoration={markings,
			mark=at position 0.5 with {\arrow[line width=1pt]{>}}}] (-1.0, 0) -- (0.0, 0.0);
		\node at (-0.5, 0.2) {\textcolor{green2}{$\mathcal{C}_2$}};
		\filldraw[black] (0, 0) circle (.03cm);
	\end{tikzpicture}
	\caption{The time contour of the numerical integration\protect\footnotemark. We evolve in Euclidean time first and then in Lorentzian time. When $|\eta|>|\eta_*|$, the conformal time is rotated by $90^\circ$, corresponding to the \textcolor{blue2}{$\mathcal{C}_1$} path.  When $|\eta|<|\eta_*|$, the integration is performed along the real axis, following the \textcolor{green2}{$\mathcal{C}_2$} path. By Cauchy's theorem, this deformation is equivalent to the original contour, and the arc at infinity can be neglected thanks to the choice of the Bunch-Davies vacuum. See \cite{Cabass:2022rhr,Stefanyszyn:2023qov} for more details.}  \label{Figure: Integration_Contour}
\end{figure}
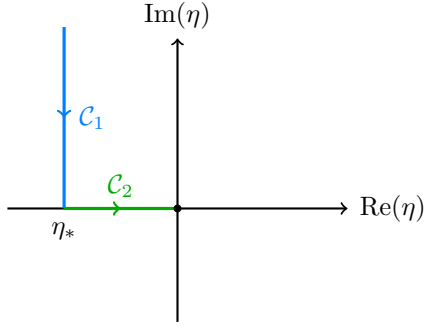
To triple check the scalar-tensor CCF formula, we also perform a direct numerical integration at the level of \eqref{B4_SKexpression}. In Figure \ref{Fig:Num}, we compare our CCF formula \eqref{CSCCF} with the numerical results. The numerical data (gray dots) are obtained via a brute-force evaluation of the SK integral in  (\ref{B4_SKexpression}). Although the numerical integrals are seemingly straightforward as the integrand consists of elementary functions only, each term in (\ref{B4_SKexpression}) is, as we have emphasised above, individually IR-divergent. Note also that the cancellation of these divergences are manifest only when the integrand is written in Lorentz time. Yet in the Lorentz signature, the integrand is highly oscillatory in the UV region as the mode functions behave like $\sim e^{\pm i k\eta}$, which must, of course, be regulated by the $i\epsilon$-prescription. One might naively think that we can perform a complete wick rotation with the $i\epsilon$-prescription, to the Euclidean time ($\eta=\pm i\chi$) so that integrand becomes rapidly decaying in the UV. However, it turns out that a complete wick rotation jeopardizes the cancellation of IR divergences, due to the different rotational directions for time-ordered (with $1-i\epsilon$) and anti-time-ordered (with $1+i\epsilon$) contours. Therefore, to regulate the UV oscillations while preserving the fine-tuned IR behaviour, we perform an incomplete Wick rotation which corresponds to adopting an ``${\sf{L}}$"-shaped contour in the numerical integration, as shown in Figure \ref{Figure: Integration_Contour}. Take a time-ordered vertex for instance, we deform the time contour as 
\begin{align}
    \eta=\eta_*+i \chi,\quad\text{if}\quad|\eta|>|\eta_*|\,,
\end{align}
whereas if $|\eta|<|\eta_*|$, we leave the integrand unchanged in its original Lorentzian signature. The choice of the turning point $\eta_*$ is arbitrary and does not affect the final result. In our numerical codes we set $\eta_*=-1/2$. As expected, this deformed contour yields rapid numerical convergence, avoids IR divergence, and agrees well with the analytical expression.

\footnotetext{This is just an illustration for the time-ordered vertices. For the anti-time-ordered ones, one should flip the ``${\sf{L}}$"-shaped contour to a ``$\sf \Gamma$''-shape one, rotating time variables counter-clockwise by $90^{\circ}$ and starting the contour from the lower half-plane.} 

\begin{figure}[h!]
    \centering
    \hspace{0.5cm}
    \begin{overpic}[width=1\linewidth]{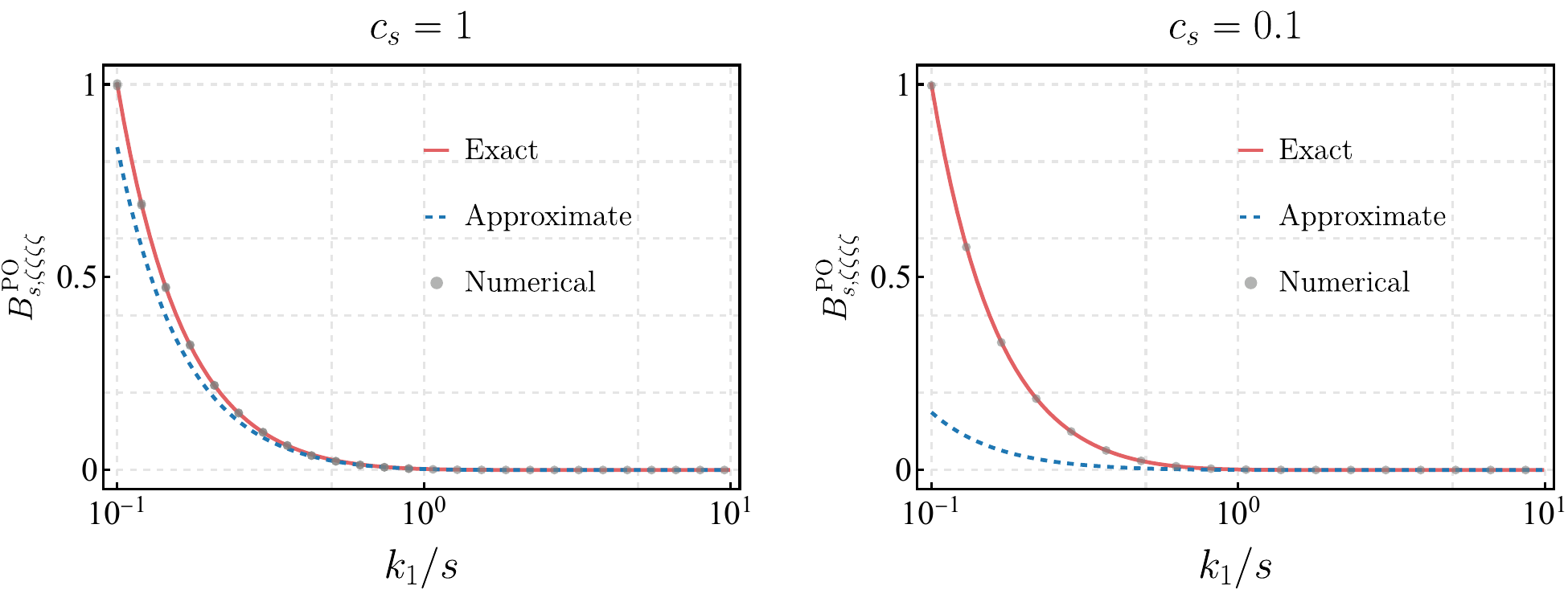}
    \put(40,23.7){\cite{Creque-Sarbinowski:2023wmb}}
    \put(91.7,23.7){\cite{Creque-Sarbinowski:2023wmb}}
    \end{overpic}
    \caption{The comparison of different results. The \textcolor[RGB]{226,97,100}{red solid line} is the  \textit{exact} result from our CCF formula \eqref{curvatureTrispectrum}, the \textcolor[RGB]{31,119,180}{blue dashed line} shows the \textit{approximate} result from the previous work \cite{Creque-Sarbinowski:2023wmb}
    and \textcolor[RGB]{178,179,178}{gray dots} represent the  \textit{numerical} results. {\textit{Left} and \textit{Right} panels correspond  to sound speeds $c_s=1$ and $c_s=0.1$, respectively}. For better visualisation, the overall size is normalised to its maximum value. We set the tetrahedron kinematics configuration as $k_1=k_3,\,k_2=k_4=\sqrt{s^2+k_1^2}$, and $\psi=\pi/3$ being the dihedral angle from $(\bfk_1,\bfk_2)$-plane to $(\bfk_3,\bfk_4)$-plane.}\label{Fig:Comparison}
\end{figure}
In addition to comparing our exact expression derived from the scalar-tensor CCF formula with brute-force numerical calculations, we also compare it with the approximate result from the previous work \cite{Creque-Sarbinowski:2023wmb}, as shown in Figure \ref{Fig:Comparison}. The two results for the trispectrum differ significantly in terms of analytical expressions. Indeed, our exact expression is factorised and is therefore without any total-energy poles, whereas the approximate result of \cite{Creque-Sarbinowski:2023wmb} is not factorised and does indeed have total-energy poles. In general, the expression found in \cite{Creque-Sarbinowski:2023wmb} is somewhat more complex and lengthy than our result and at the level of these expressions there is no reason to believe that one is an approximation of the other. Despite the dramatically different structures in the analytical expressions, the resulting shapes deviate by no more than $20\%$ as can be seen in the left panel of Figure \ref{Fig:Comparison}. Thus one might be led to conclude that the result in \cite{Creque-Sarbinowski:2023wmb} is always an accurate tracer of our exact result. However, we note that this resemblance is accidental and disappears when we turn on a non-unit sound speed of the inflaton. In the EFT of inflation, the sound speed of the inflaton $c_s$ can be reduced via the inclusion of higher-dimensional operators \cite{Cheung:2007st}. At the correlator level, this amounts to rescaling the external energies by $k_{\sf a}\to c_s k_{\sf a}, \sf a=1,2,3,4$, with $0<c_s<1$. Performing this rescaling in the scalar trispectrum, we observe our exact result always agrees with the numerics well and can be significantly larger than the approximate prediction of \cite{Creque-Sarbinowski:2023wmb}, as shown in the right panel of Figure \ref{Fig:Comparison}. We find this quite interesting and believe that it deserves further investigation which we leave to future work. 

\section{From wavefunctions to correlators: a parity-odd subtlety}
In this appendix we provide more details on how we arrive at \eqref{WavefunctionToCorrelator} which tells us how we go from wavefunction coefficients to cosmological correlators. The derivation is somewhat simple, we are required to expand around the Gaussian of \eqref{CorrelatorAverage} and perform various Gaussian integrals. The purpose of this appendix is to be careful with minus signs and show why the arguments on the RHS on \eqref{WavefunctionToCorrelator} pick up minus signs. For parity-even theories such a subtlety is not an issue, but for parity-odd correlators it is important so that we get the overall sign of the final correlator correct. 

We will concentrate on the four-point function of inflatons and so we want to compute 
\begin{align} \label{CorrelatorAverage}
\langle \bar{\varphi}(\bfk_{1}) \ldots \bar{\varphi}(\bfk_{4})  \rangle  = \frac{\int \mathcal{D} \bar{\varphi} \mathcal{D} \bar{\gamma}\, |\Psi|^2 \,  \bar{\varphi}(\bfk_{1}) \ldots \bar{\varphi}(\bfk_{4})}{\int \mathcal{D} \bar{\varphi} \mathcal{D} \bar{\gamma}\, |\Psi|^2} \,.
\end{align}
For illustrative purposes we take the wavefunction to be parametrised by
\begin{align}
\Psi = \exp[-\frac{1}{2} \int_{\bfk }\psi_{\varphi \varphi}
(k)  \bar{\varphi}(-\bfk) \bar{\varphi}(\bfk) + \frac{1}{24} \int_{\bfk_1 \ldots \bfk_4}  \delta \psi_{\varphi \varphi \varphi \varphi } (\{ \bfk \}) \hat{\delta}\left(\sum \bfk \right)   \bar{\varphi}(\bfk_1) \ldots \bar{\varphi}(\bfk_4)]\,, \label{WavefunctionParametrisation} \\
|\Psi|^2 = \exp[-\frac{1}{2} \int_{\bfk }\rho_{\varphi \varphi}
(k)  \bar{\varphi}(-\bfk) \bar{\varphi}(\bfk) + \frac{1}{24} \int_{\bfk_1 \ldots \bfk_4}  \delta \rho_{\varphi \varphi \varphi \varphi } (\{ \bfk \}) \hat{\delta}\left(\sum \bfk \right)   \bar{\varphi}(\bfk_1) \ldots \bar{\varphi}(\bfk_4)]\,,\label{WavefunctionSquaredParametrisation}
\end{align}
where $\rho_{\varphi \varphi}
(k) = \psi_{\varphi \varphi}
(k)+\psi^{*}_{\varphi \varphi}
(k)$ and $\delta \rho_{\varphi \varphi \varphi \varphi } (\{ \bfk \}) = \delta \psi_{\varphi \varphi \varphi \varphi } (\{ \bfk \})+\delta \psi^*_{\varphi \varphi \varphi \varphi } (\{ -\bfk \})$. To go from \eqref{WavefunctionParametrisation} to \eqref{WavefunctionSquaredParametrisation} we need to use $\bar{\varphi}(\bfk) = \bar{\varphi}^*(-\bfk)$ which follows from the reality of the field in position space. This wavefunction is a subset of the full \eqref{WavefunctionExpansion}. The other terms do not change the qualitative picture of what comes next and can be added with little extra effort. Since we are interested in the connected four-point function at tree-level we can expand the exponentials such that 
\begin{align} 
\langle \bar{\varphi}(\bfk_{1}) \ldots \bar{\varphi}(\bfk_{4})  \rangle  = \frac{\int \mathcal{D} \bar{\varphi}\, e^{-\frac{1}{2} \int_{\bfk }\rho_{\varphi \varphi}
(k)  \bar{\varphi}(-\bfk) \bar{\varphi}(\bfk)} \int_{\bfq_1 \ldots \bfq_4}  \delta \rho_{\varphi \varphi \varphi \varphi } (\{ \bfq \}) \hat{\delta}\left(\sum \bfq \right)   \bar{\varphi}(\bfq_1) \ldots \bar{\varphi}(\bfq_4)\times  \bar{\varphi}(\bfk_{1}) \ldots \bar{\varphi}(\bfk_{4})}{24 \int \mathcal{D} \bar{\varphi} \, e^{-\frac{1}{2} \int_{\bfk }\rho_{\varphi \varphi}
(k)  \bar{\varphi}(-\bfk) \bar{\varphi}(\bfk)}} \,.
\end{align}
We now want to take the real and imaginary parts of $\bar{\varphi}$ as independent variables with 
\begin{align}
\bar{\varphi}(\bfk) = \text{Re}\bar{\varphi}(\bfk) + i \text{Im}\bar{\varphi}(\bfk) \,, \\ 
\bar{\varphi}(-\bfk) = \text{Re}\bar{\varphi}(\bfk) - i \text{Im}\bar{\varphi}(\bfk) \,, 
\end{align}
where again we have used $\bar{\varphi}(\bfk) = \bar{\varphi}^*(-\bfk)$. The same relations hold for the graviton meaning that including the graviton is a simple extension of what we are doing here. We can therefore write 
\begin{align}
e^{-\frac{1}{2} \int_{\bfk }\rho_{\varphi \varphi}
(k)  \bar{\varphi}(-\bfk) \bar{\varphi}(\bfk)} = e^{-\frac{1}{2} \int_{\bfk }\rho_{\varphi \varphi}
(k)[\text{Re}\bar{\varphi}(\bfk)^2+\text{Im}\bar{\varphi}(\bfk)^2]}\,,
\end{align}
and consider separate Gaussian integrals over $\text{Re}\bar{\varphi}$ and $\text{Im}\bar{\varphi}$. In terms of these variables we have
\begin{align}
\bar{\varphi}(\bfq_1) \ldots \bar{\varphi}(\bfq_4)  \bar{\varphi}(\bfk_{1}) \ldots \bar{\varphi}(\bfk_{4}) =&[(\text{Re}\bar{\varphi}(\bfk_1) + i \text{Im}\bar{\varphi}(\bfk_1)) \ldots (\text{Re}\bar{\varphi}(\bfk_4) + i \text{Im}\bar{\varphi}(\bfk_4)) ] \nonumber \\
\times &[(\text{Re}\bar{\varphi}(\bfq_1) + i \text{Im}\bar{\varphi}(\bfq_1)) \ldots (\text{Re}\bar{\varphi}(\bfq_4) + i \text{Im}\bar{\varphi}(\bfq_4)) ] \,.
\end{align}
Since we are computing the connected four-point function, we are interested in terms where the delta function $\hat{\delta}\left(\sum \bfq \right)$ becomes $\hat{\delta}\left(\sum \bfk \right)$ after we have performed all integrations. For this to occur, we need to pair the $\bfq$'s and $\bfk$'s with a pair defined by a $\bfq$ being equal or opposite to a $\bfk$. There are 24 possible sets of pairs, with each set containing 4 pairs. In a given set the signs must be uniform. For example, for the set $(\bfq_1, \bfk_1), \ldots ,(\bfq_4, \bfk_4)$, we have two choices $(\bfq_1 = \pm \bfk_1), \ldots ,(\bfq_4 = \pm \bfk_4)$. The first of these choices i.e. $(\bfq_1 = \bfk_1), \ldots ,(\bfq_4 = \bfk_4)$ yields
\begin{align}
\bar{\varphi}(\bfq_1) \ldots \bar{\varphi}(\bfq_4)  \bar{\varphi}(\bfk_{1}) \ldots \bar{\varphi}(\bfk_{4}) =&[(\text{Re}\bar{\varphi}(\bfk_1) + i \text{Im}\bar{\varphi}(\bfk_1)) \ldots (\text{Re}\bar{\varphi}(\bfk_4) + i \text{Im}\bar{\varphi}(\bfk_4)) ] \nonumber \\
\times &[(\text{Re}\bar{\varphi}(\bfk_1) + i \text{Im}\bar{\varphi}(\bfk_1)) \ldots (\text{Re}\bar{\varphi}(\bfk_4) + i \text{Im}\bar{\varphi}(\bfk_4)) ] \nonumber \\
=&[\text{Re}\bar{\varphi}(\bfk_1)^2 -  \text{Im}\bar{\varphi}(\bfk_1)^2] \ldots [(\text{Re}\bar{\varphi}(\bfk_4)^2 - \text{Im}\bar{\varphi}(\bfk_4)^2] \nonumber \\
& + \text{odd terms} \,,
\end{align}
where we have not written out terms that are odd since they yield a vanishing result when integrated over. If we now perform the Gaussian integrals it is clear that the contribution from $\text{Re}\bar{\varphi}(\bfk_1)^2$ will cancel with that from $\text{Im}\bar{\varphi}(\bfk_1)^2$ due to the relative minus sign. The same is true for the other momenta, and for any other set where the $\bfq$'s and $\bfk$'s are equal, and so for this choice there is no contribution to the connected four-point function. The second choice i.e. $(\bfq_1 = - \bfk_1), \ldots ,(\bfq_4 = - \bfk_4)$ instead yields 
\begin{align}
\bar{\varphi}(\bfq_1) \ldots \bar{\varphi}(\bfq_4)  \bar{\varphi}(\bfk_{1}) \ldots \bar{\varphi}(\bfk_{4}) =&[(\text{Re}\bar{\varphi}(\bfk_1) + i \text{Im}\bar{\varphi}(\bfk_1)) \ldots (\text{Re}\bar{\varphi}(\bfk_4) + i \text{Im}\bar{\varphi}(\bfk_4)) ] \nonumber \\
\times &[(\text{Re}\bar{\varphi}(\bfk_1) - i \text{Im}\bar{\varphi}(\bfk_1)) \ldots (\text{Re}\bar{\varphi}(\bfk_4) - i \text{Im}\bar{\varphi}(\bfk_4)) ] \nonumber \\
=&[\text{Re}\bar{\varphi}(\bfk_1)^2 +  \text{Im}\bar{\varphi}(\bfk_1)^2] \ldots [(\text{Re}\bar{\varphi}(\bfk_4)^2 + \text{Im}\bar{\varphi}(\bfk_4)^2] \nonumber \\
& + \text{odd terms} \,,
\end{align}
where now some crucial signs have changed. The contributions from $\text{Re}\bar{\varphi}(\bfk_1)^2$ and $\text{Im}\bar{\varphi}(\bfk_1)^2$ no longer cancel and we find a non-zero result.\footnote{The analogous statement in the in-in formalism is that the non-zero Wick contractions are only between pairs of fields with equal and opposite momenta.} If we now go ahead and compute the Gaussian integrals, including all possible sets of pairs, we find \eqref{WavefunctionToCorrelator} (when we include the additional terms in the wavefunction). We therefore see why on the RHS of \eqref{WavefunctionToCorrelator} we have e.g. $\delta \rho_{\varphi \varphi \varphi \varphi}(- \bfk_1, -\bfk_2, -\bfk_3, -\bfk_4)$ rather than $\delta \rho_{\varphi \varphi \varphi \varphi}( \bfk_1, \bfk_2, \bfk_3, \bfk_4)$. 
\newpage

\bibliographystyle{JHEP}
\bibliography{refs}

\end{document}